\documentclass[preprint]{aastex}
\usepackage{lscape}
\usepackage{longtable}
\usepackage{amssymb,latexsym,amsmath}
\usepackage{graphicx}

\shorttitle{A tail of two tails}
\shortauthors{Marelli et al.}

\begin{document}

\title{The tale of the two tails of the oldish PSR J2055+2539}

\author{Martino Marelli\altaffilmark{1}, Daniele Pizzocaro\altaffilmark{1,2}, Andrea De Luca\altaffilmark{1}, Fabio Gastaldello\altaffilmark{1}, Patrizia Caraveo\altaffilmark{1},
Pablo Saz Parkinson\altaffilmark{3,4}}

\altaffiltext{1}{INAF - Istituto di Astrofisica Spaziale e Fisica Cosmica Milano, via E. Bassini 15, 20133 Milano, Italy}
\altaffiltext{2}{Universit\'a degli Studi dell'Insubria - Via Ravasi 2 - 21100 Varese - Italy}
\altaffiltext{3}{Department of Physics, The University of Hong Kong, Pokfulam Road, Hong Kong}
\altaffiltext{4}{Santa Cruz Institute for Particle Physics, University of California, Santa Cruz, CA 95064}

\begin{abstract}
We analyzed a deep {\it XMM-Newton} observation of the radio-quiet $\gamma$-ray PSR J2055+2539.
The spectrum of the X-ray counterpart is non-thermal, with a photon index of $\Gamma$=2.36$\pm$0.14 (1$\sigma$ confidence). We detected X-ray pulsations
with a pulsed fraction of (25$\pm$3)\% and a sinusoidal shape.
Taking into account considerations on the $\gamma$-ray efficiency of the pulsar and on its X-ray spectrum,
we can infer a pulsar distance ranging from 450 pc to 750 pc.
We found two different nebular features associated to PSR J2055+2539 and protruding from it. The angle between the two nebular main axes is
$\sim$ (162.8$\pm$0.7)$^{\circ}$.
The main, brighter feature is 12' long and $<$20" thick, characterized by
an asymmetry with respect to the main axis that evolves with the distance from the pulsar, possibly forming a helical pattern.
The secondary feature is 250" $\times$ 30".
Both nebulae present an almost flat brightness profile with a sudden decrease at the end.
The nebulae can be fitted either by a power-law model or a thermal bremsstrahlung model.
A plausible interpretation of the brighter nebula is in terms of a collimated ballistic jet.
The secondary nebula is most likely a classical synchrotron-emitting tail.
\end{abstract}


\section{Introduction} \label{intro}

The Large Area Telescope on board the {\it Fermi} satellite \citep[Fermi-LAT][]{atw09} has opened a new era for pulsar astronomy by detecting $\gamma$-ray pulsations
(at E $>$ 100 MeV) from more than 160 pulsars \footnote{\url{https://confluence.slac.stanford.edu/display/GLAMCOG/Public+List+of+LAT-Detected+Gamma-Ray+Pulsars}},
$\sim$ 25\% of which are not detected at radio wavelengths. The discovery of such a large sample of high-energy-emitting pulsars
has enabled numerous population studies \citep[see e.g.][]{wat11,pie12}.
This has resulted in a better understanding of the physical phenomena behind the multiwavelength emission from pulsars.

Some of the {\it Fermi} pulsars deserved dedicated multiwavelength campaigns. Such ``extreme'' objects, accounting for the tails of
the population distribution in energetics, age and magnetic field, are key to our understanding of the entire population, posing important limits on
the different pulsar emission models. These campaigns have sometimes resulted in the discovery of important and unexpected features: e.g. the peculiar multiwavelength light curve profile of the energetic
radio-quiet pulsar PSR J1813-1246 disagrees with the classical model for X-ray emission of pulsars \citep{mar14b,kui15}; the isolated PSR J2021+4026 shows hints of long-term variations
in the $\gamma$-ray flux, challenging current high-energy emission models \citep{all13}.

The increase in the number of $\gamma$-ray pulsars has been crucial to the study of phenomena related to highly energetic pulsars, such as X-ray and radio nebulae.
Such features are usually interpreted
within the framework of bow-shock, ram-pressure-dominated, pulsar wind nebulae \citep[PWNe; for a review see][]{gae06}.
If the pulsar moves supersonically, shocked pulsar wind is expected to flow in an elongated region downstream of
the termination shock (basically, the cavity in the interstellar medium (ISM) created by the moving neutron star and its
wind), confined by ram pressure. X-ray emission is due to synchrotron emission from the wind particles accelerated at
the termination shock, which is typically seen (if angular resolution permits) as the brightest portion of the extended
structure \citep[see e.g.][]{kar08a}, as predicted by MHD simulations \citep{buc02,van03,buc05}.
In order to produce a population of high-energy particles, a higly-energetic pulsar is
required \citep[$\dot{E}$ greater than $\sim$ 10$^{34}$, see e.g.][]{kar08b}.
For classical synchrotron nebulae, we expect a relatively bright diffuse emission surrounding the pulsar, where the emission from the
wind termination shock is brightest, as observed in all the other known cases \citep[see e.g.][]{gae04,mcg06,kar08b}.
The brightness is only slightly dependent on the interstellar medium density ($\propto$ n$_{ISM}^{1/2}$),
thus we expect a relatively uniform brightness profile.
One of the newly-discovered pulsars, PSR J0357+3205 \citep{abd09}, revealed a bright tail that cannot be interpreted in the framework of synchrotron PWNe,
due to the low energetics of the leading pulsar and to a peculiar shape \citep{del11}.
Alternatively, a new model for pulsar nebulae was proposed, involving bremsstrahlung thermal emission from the ISM, shocked by the supersonically-moving pulsar,
accounting for all the characteristics of the nebula \citep{mar13}. The cooldown time for this emission is orders of magnitude longer than for the synchrotron model,
thus implying a lack of spectral variation along the tail. Pulsar bremsstrahlung nebulae (PBN) emission is much more dependent on the ISM density ($\propto$ n$_{ISM}^2$),
thus pointing to a clumpy brightness profile, with the brightness following and enhancing the medium distribution map. For this type of nebula we expect that
most of the energy in the pre-shock flow is carried by the ions, while the electron temperature is responsible for the X-ray emission: we do not expect X-ray emission
near the pulsar, due to the Coulomb heating time of electrons. Both PWNe and PBNe obviously require the proper motion of the pulsar to be aligned with the main axis nebular emission,
in the case of a cometary shape.
Recently, a peculiar nebular feature was found around the pulsar IGR J11014-6103 \citep{pav11}. A deep {\it Chandra} observation revealed a helical structure protruding from the point source,
possibly associated with a jet-like emission. A smaller trail was detected aligned with the direction of the putative, very large ($>$ 10$^3$ km s$^{-1}$) proper motion, and it was associated to a classical pulsar wind nebular emission. This is reminiscent of the jet of the Guitar Nebula \citep[PSR B2224+65,][]{hui07}.
This feature is usually explained in terms of a collimated ballistic jet, even if more complex explanations have been developed \citep{ban08}.
The helical shape and the collimated structure are the distinctive trait of this type of pulsar jet nebula (PJN). Another fundamental characteristic is
the misalignment with the direction of proper motion.

Among the extremes of the population of {\it Fermi} pulsars, PSR J2055+2539 \citep[J2055 hereafter,][]{saz10}
is one of the best candidates. It is one of the 100 brightest $\gamma$-ray sources, with a $\gamma$-ray
flux of (5.5$\pm$0.1) $\times$ 10$^{-7}$ erg cm$^{-2}$ s$^{-1}$ \citep{ace15}. It is located far from the Galactic plane, at a latitude $\sim$-13$^{\circ}$, making the search for
counterparts at other wavelengths easier. A blind search using {\it Fermi} data led to the unambiguous detection of the timing signature of a pulsar with P $\sim$ 320 ms and $\dot{P}$ $\sim$ 4.1 $\times$ $10^{-14}$ s s$^{-1}$ \citep{abd13}.
J2055 is among the less energetic and oldest pulsars in the {\it Fermi} zoo, with a spin-down energy of $\dot{E}$ = 5.0 $\times$ 10$^{33}$ erg s$^{-1}$ and a characteristic age
$\tau_c$ = 1.2 Myr. The $\gamma$-ray ``pseudo-distance'' relation \citep{saz10}, based on an empirical relation among $\gamma$-ray pulsar luminosities and their
spin-down energies, yields an estimated distance of 600 pc.
The second {\it Fermi} pulsar catalog \citep{abd13} reports J2055 to be one of the very few $\gamma$-ray pulsars with a strong detection of off-pulse emission, characterized by a spectral cut-off: this emission probably originates from its magnetosphere, opening a new conundrum for magnetospheric emission models. In fact, for young pulsars this is a serious
challenge to outer-magnetosphere models radiating only above the null-charge surface; such weakly pulsed emission should be rare, being expected only for nearly-aligned pulsars.

Due to its peculiar characteristics, after two short {\it XMM-Newton} exploratory observations in 2013 we obtained a long, uninterrupted XMM-Newton observation
to study the X-ray counterpart of J2055 and to search for X-ray pulsations.
Section \ref{obs} reports the characteristics of the observations and the analysis methods. In Section \ref{field} we report the analysis of the serendipitous sources in the field.
Section \ref{pulsar} describes the spectral and timing characteristics of the pulsar in the X-ray band.
This campaign revealed the presence of spectacular nebular features, described in Section \ref{nebula}. Pulsar and nebular modeling are treated in sections \ref{disc} and \ref{conc}.
In the appendices, we describe the details of the field analysis (Appendix \ref{appa}) and of the new script we developed for our {\it XMM-Newton} analysis (Appendix \ref{appb}).

\section{Observations and Data Reduction} \label{obs}

Our deep {\it XMM-Newton} observation of J2055, lasting 136.2 ks, was performed on 2013 May 1 (obs. id 0724090101). The PN camera \citep{str01} of the EPIC instrument was operating
in Large Window mode, with a time resolution of 47.7 ms on a 27'$\times$13' field of view (FOV). The high time resolution combined
with the large FOV allow for both the timing analysis of the J2055 pulsar and the spatial analysis of the nebular structures.
The Metal Oxide Semi-conductor (MOS) detectors \citep{tur01} were set in Full Frame mode (2.6 s time resolution on
a 15' radius FOV). The thin optical filter was used both for PN and MOSs.
We also analyzed XMM-Newton observations 0605470401 and 0605470901, taken on 2009 October 26 and on 2010 April 21, and lasting 24.5
and 17.9 ks respectively. The PN camera was operating in full frame mode,
with a time resolution (73.4 ms). A timing analysis for this pulsar would be possible even using full-frame PN data, but it would
result in a lower resolution of the light curve. We therefore used data from all the observations
for the following analysis, with the exception of the timing analysis for which we considered only PN data of the longest observation.

\begin{figure}
\epsscale{1.0}
\plotone{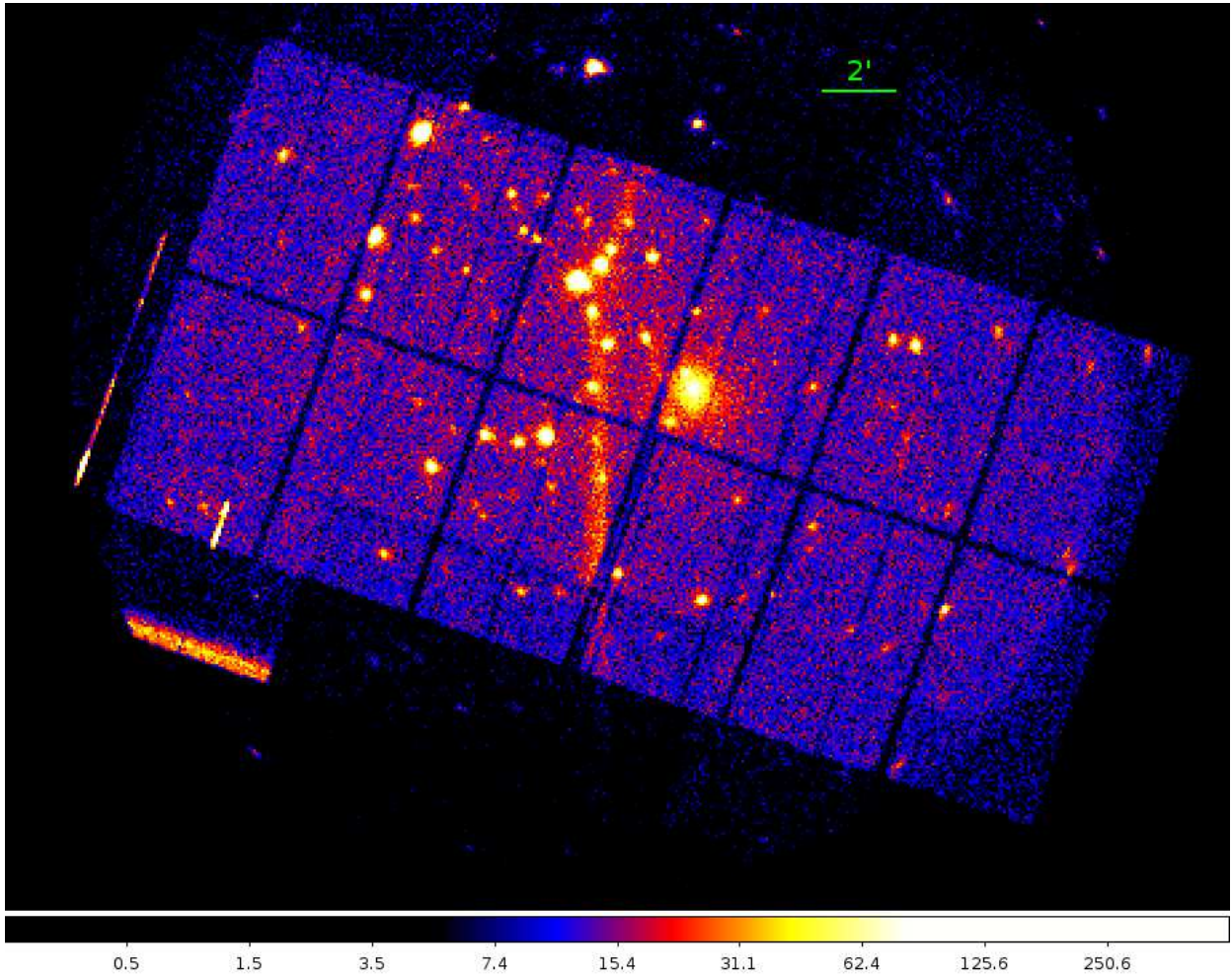}
\caption{0.3-10 keV sky maps showing only data used for the analysis in this paper: this includes the pattern, flag and proton flares filters described in Sections \ref{obs}, \ref{field}.
PN and MOSs cameras images from the three observations were summed.
\label{fig-excl}}
\end{figure}

We used the {\it XMM-Newton} Science Analysis Software (SAS) v13.5.
For the standard analysis, a tool included in the SAS {\tt epproc} usually randomizes the arrival time of each event within the time resolution.
Here, we chose not to randomize the arrival times in order to achieve a better timing resolution for our pulsar.
A discussion on the implication of this choice is reported in \citet{mar14b}. We performed a standard analysis of
high particle background with the SAS tool {\tt bkgoptrate} (also used for the compilation of the 3XMM source
catalog \footnote{\url{http://xmmssc.irap.omp.eu/Catalogue/3XMM-DR5/3XMM\_DR5.html}}). This
tool searches for the point at which the maximum signal-to-noise (S/N)
ratio is achieved for the given background time series: time bins above this threshold are excluded.
We then cross-checked the results using an independent method, following \citet{del05}.
Both the analyses revealed a high contamination from flares during almost the entire observation 0605470901, that was therefore excluded from subsequent analysis.
Moderate contamination was detected in the other observations, lowering the good time interval to 132 ks.
Following the prescription from the 3XMM source catalog, for the PN camera we selected 0-4 pattern events in the 0.5-10 keV energy range
and 0 pattern events in the 0.3-0.5 keV energy range. We applied standard flags to remove bright columns.
We selected 0-12 pattern events for the MOS detectors in the 0.3-10 keV energy range, applying a similar reduction for bright columns.
Finally, we excluded out-of-FOV events and residual contaminants.
Figure \ref{fig-excl} shows images of the counts satisfying our event selection criteria from all the datasets.
For the timing analysis, we used the SAS tool {\tt barycen} to barycenter the PN events at the best-fit XMM-Newton pulsar position (see Section \ref{field}).
For each spectrum, we generated ad-hoc response matrices and effective area files using the SAS tools {\tt rmfgen} and {\tt arfgen}.\\
We checked for other X-ray observation of the source and found a {\it Suzaku} \citep{mit07} observation (obs. id 405015010) taken on 2010 October 29
lasting 31.1 ks. The faintness of the X-ray counterpart, the crowding of the field (see Appendix \ref{appa}), coupled with the low spatial
resolution of Suzaku (the half-power diameter of the XIS instrument, onboard {\it Suzaku}, is $\sim$ 2$'$) make this observation useless for our purposes.
It was thus excluded from our analysis.

\section{Field Analysis} \label{field}

\begin{figure}
\epsscale{1.0}
\plotone{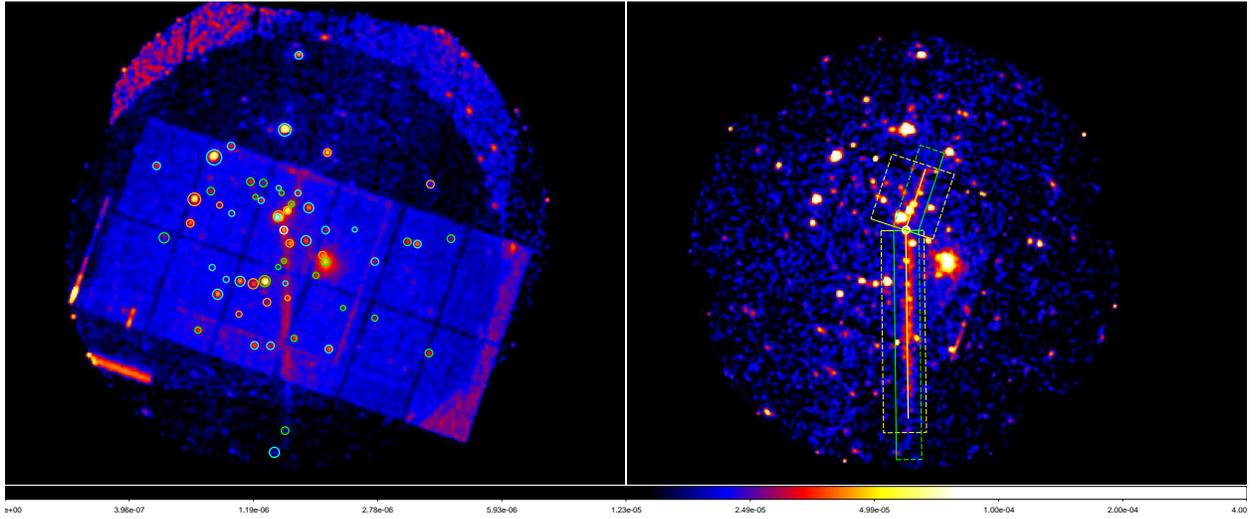}
\caption{Exposure-corrected sky maps showing events from all the considered datasets (left, see Section \ref{obs})
and only from the MOS2 datasets of the deep {\it XMM-Newton} observation (right).
While the first image shows the entire dataset we used, the larger FOV of the MOS2 camera allows for a better view of the nebulae.
Events have been extracted as described in Section \ref{obs}; the images are in logarithmic scale. {\it Left panel}: We reported the extraction regions used for
serendipitous sources analysis, with a color code showing the most probable association. White: the pulsar; cyan: AGN association ; orange: star association;
green: no association. {\it Right panel}: We reported the best-fit main axes of the nebular features together with the best-fit length (yellow lines).
We also indicate the areas used for the spatial analysis of the nebulae: $''$AMA$''$ in green and $''$OMA$''$ in yellow (see Section \ref{nebula}).
\label{fig-field}}
\end{figure}

Figure \ref{fig-field} shows the 0.3-10 keV {\it XMM-Newton} FOV.
Source detection using maximum likelihood fitting was done both
simultaneously and on each of the EPIC-PN, MOS1, and MOS2 datasets, using the SAS
tool {\tt edetect\_chain}. The resulting detections were then confirmed using the CIAO tool {\tt wavdetect},
following prescriptions from the CIAO online cookbook \footnote{\url{http://cxc.harvard.edu/ciao/threads/all.html}}.

The best position of the pulsar X-ray counterpart is 20$^h$55$^m$48$^s$.89 +25$^{\circ}$39$'$57.8$''$ (0.3$''$+1.5$''$ 1$\sigma$ statistical plus systematic errors).
From the most recent compilation of the LAT $\gamma$-ray
Pulsar Timing Models \footnote{\url{https://confluence.slac.stanford.edu/display/GLAMCOG/LAT+Gamma-ray+Pulsar+Timing+Models}},
the best $\gamma$-ray position of the pulsar is 20$^h$55$^m$48$^s$.94 +25$^{\circ}$39$'$59.1$''$  (0.5$''$ 1$\sigma$ statistical plus systematic error), fully compatible with the X-ray position.
According to the logN-logS distribution of the {\it Chandra} galactic sources \citep{nov09} at mid Galactic latitudes,
we can estimate the probability of a chance detection of an
X-ray source within the LAT $\gamma$-ray timing error box and with X-ray flux
comparable or greater than the X-ray counterpart ($\sim$ $10^{-14}$ erg cm$^{-2}$ s$^{-1}$) to be less than 10$^{-5}$. Thus, based on positional
coincidence, we consider our identification secure.

In Figure \ref{fig-field} we marked the brightest sources in the PN field of view.
A bright nebular source is apparent in the field. This object, possibly associated with a galaxy cluster,
will be described in Gastaldello et al. (in preparation) and not further described in this paper.
We chose all the sources with an absorbed flux greater than 5 $\times$ $10^{-15}$ erg cm$^{-2}$ s$^{-1}$,
with a 5$\sigma$ detection and with at least 150 {\it XMM-Newton} net, total counts.
We also considered all the sources detected in the nebular regions.
The study of the line-of-sight absorption of such sources could allow us to constrain the
pulsar distance. Indeed, after selecting candidate Active Galactic Nuclei 
(AGN) in the FOV, it is possible to measure, from their spectra,
the total Galactic column density 
in the direction of J2055. Next, using the pulsar column density, we can get 
an estimate of its distance with respect to the
edge of the Galaxy. Such estimate could be refined if bright X-ray-emitting 
and optical-emitting stars (with known distances) were also present
\citep[see e.g.][]{mar13,mar14a,mar14b}.
The spatial resolution of this measurement is much higher than the one from the 21 cm H$_{\hbox{\scriptsize I}}$ sky survey of \citet{kal05}.
The resolution of the survey is 0.6$^{\circ}$, larger than the entire {\it XMM-Newton} FOV:
the presence of local molecular clouds could affect the H$_{\hbox{\scriptsize I}}$ result even by orders of magnitude \citep[see e.g. the case of PSR J1813-1246 field][]{mar14b}.

Based on the study of the spectra of serendipitous sources and of associated optical counterparts,
we classified serendipitous sources as AGNs or candidate stars (see Appendix \ref{appa}). 
Our exercise allowed us to identify 24 very likely AGN and 13 stars from a total
of 58 X-ray sources. In particular, one of the brightest sources (source 1 in Table 1)
was originally detected as a single source: only through the application of a new script were we able to determine that it consisted of
the superimposition of multiple sources (see Appendix \ref{appb} for details).

The simultaneous fit of AGN spectra, with the column density linked, results in a relatively low value of column density: (2.07$\pm$0.08) $\times10^{21}$ cm$^{-2}$ (1$\sigma$ error)
as expected in a mid-latitude region. This result is higher than
the value of 1.2 $\times$ 10$^{21}$ cm$^{-2}$ obtained from the 21 cm H$_{\hbox{\scriptsize I}}$ sky survey of \citet{kal05}.

\setlength{\LTleft}{-1pt}
\begin{footnotesize}
\begin{landscape}
\begin{center}
\begin{longtable}{cccccc}
source & J2000 coord & spectrum & n$_H$ & $log(\frac{f_{X}}{f_{opt}})^a$ & Id\\
- & - & - & $10^{21}\,cm^{-2}$ & - & -\\
\endhead
1 & 313.96143 25.679886 & apec+pow/2apec & 4.77$_{-1.21}^{+0.78}$/4.00$_{-0.27}^{+0.53}$ & $>$0.566+0.686/$>$1.12(V*) & star+agn\\
2 & 313.94955 25.687331 & 2apec & 0.981$_{-0.103}^{+0.118}$ & -1.58(V) & star\\
3 & 313.94435 25.694125 & pow & 65.5$_{-11.0}^{+12.8}$ & ? & agn$^b$\\
4 & 313.93596 25.706747 & pow & 1.23$_{-0.48}^{+0.60}$ & $>$-0.291(V*) & agn\\
6 & 313.94593 25.651206 & pow/apec & 2.79$_{-0.36}^{+0.39}$/2.06$\pm$0.25 & -0.927/-0.503(V*) & star\\
7 & 313.95363 25.631689 & pow/apec & 2.05$_{-0.48}^{+0.56}$/1.51$_{-0.28}^{+0.34}$ & $>$-0.255/$>$-0.884(V*) & ?\\
8 & 313.95196 25.607475 & pow & 2.28$_{-0.71}^{+0.86}$ & $>$-0.546(V*) & agn\\
9 & 313.9496 25.590583 & apec & 4.60$_{-0.69}^{+2.03}$ & -3.86(V) & star\\
10 & 313.94093 25.547181 & pow/apec & 2.06$_{-0.40}^{+0.44}$/1.52$_{-0.25}^{+0.26}$ & $>$-0.051/$>$0.332(V*) & ?\\
13 & 313.96099 25.6253 & pow & 0.135$_{-0.135}^{+0.298}$ & -1.86(V) & ?\\
14 & 313.97678 25.609422 & pow/2apec & 1.80$_{-0.10}^{+0.11}$/3.65$_{-0.81}^{+0.78}$ & $>$0.607/$>$0.899(V*) & ?\\
15 & 313.99079 25.606825 & pow/2apec & 1.68$_{-0.31}^{+0.35}$/2.63$_{-1.03}^{+3.29}$ & -0.387/-0.612(B) & star\\
16 & 313.97448 25.586575 & apec & 1.18$_{-0.53}^{+0.81}$ & -3.29(V) & star\\
17 & 314.00763 25.609944 & pow & 1.45$_{-0.27}^{+0.31}$ & $>$0.127(V*) & agn\\
18 & 313.96974 25.538703 & pow & 1.57$_{-0.59}^{+0.81}$ & $>$-0.367(V*) & agn\\
19 & 313.98956 25.539306 & pow & 1.33$_{-0.34}^{+0.42}$ & $>$-0.129(V*) & agn\\
20 & 313.92683 25.654367 & pow & 3.44$_{-0.50}^{+0.56}$ & $>$-0.273(V*) & agn\\
21 & 313.92368 25.690575 & pow & 2.63$_{-0.36}^{+0.40}$ & $>$-0.108(V*) & agn\\
22 & 313.98105 25.698697 & pow & 0.366$_{-0.143}^{+0.207}$ & $>$0.297(V*) & agn\\
23 & 313.98844 25.702833 & pow/apec & 5.26$_{-1.24}^{+1.56}$/4.04$_{-0.97}^{+1.50}$ & $>$-0.506/$>$-0.0326(V*) & ?\\
24 & 313.95607 25.707142 & pow/apec & 3.67$_{-1.43}^{+2.39}$/2.41$_{-0.96}^{+1.68}$ & $>$-0.695/$>$-0.166(V*) & ?\\
25 & 313.96015 25.712533 & pow & 15.6$_{-5.42}^{+6.75}$ & $>$-2.47(V*) & agn\\
26 & 313.90219 25.666081 & pow & 0.164$_{-0.08}^{+0.09}$ & $>$-0.0620(V*) & agn\\
27 & 313.86733 25.666336 & pow & 1.56$_{-1.51}^{+4.25}$ & $>$-0.310(V*) & agn\\
28 & 313.84279 25.631844 & pow & 0.816$_{-0.339}^{+0.527}$ & $>$-0.321(V*) & agn\\
29 & 313.80258 25.653275 & pow & 1.49$_{-0.33}^{+0.40}$ & -1.35(V*) & ?\\
30 & 313.79122 25.650547 & pow & 2.08$_{-0.31}^{+0.34}$ & $>$0.160(V*) & agn\\
31 & 313.74973 25.656481 & pow/apec & 8.12$_{-2.25}^{+2.93}$/7.48$_{-2.35}^{+2.57}$ & $>$-0.929/$>$-0.597(V*) & ?\\
32 & 313.88097 25.580411 & pow/apec & 3.32$_{-1.41}^{+1.89}$/2.05$_{-0.78}^{+1.14}$ & $>$-0.743/$>$-0.368(V*) & ?\\
33 & 313.84338 25.568675 & ? & ? & ? & ?\\
34 & 313.77725 25.530489 & pow & 5.50 & -1.56(V) & ?\\
35 & 313.89859 25.535194 & pow & 2.33$_{-0.43}^{+0.51}$ & $>$-0.0162(V*) & agn\\
36 & 313.91509 25.61615 & pow/2apec & 1.71$_{-0.36}^{+0.42}$/1.32$_{-0.34}^{+0.77}$ & $>$-0.192/$>$0.187(V*) & ?\\
37 & 314.00874 25.573114 & pow/apec & 4.17$_{-2.00}^{+3.23}$/4.13$_{-1.80}^{+3.78}$ & -1.66/-1.44(V*) & star\\
39 & 314.05828 25.556017 & pow/apec & 5.53$_{-1.38}^{+2.00}$/4.80$_{-1.41}^{+2.24}$ & $>$-0.540/$>$-0.147(V*) & ?\\
40 & 314.03453 25.595597 & pow & 1.32$_{-0.19}^{+0.22}$ & $>$0.176(V*) & agn\\
41 & 314.02360 25.612442 & pow & $<$0.629 & $>$-0.277(V*) & agn\\
42 & 314.04116 25.624814 & pow & 2.15$_{-0.90}^{+1.22}$ & $>$-0.653(V*) & agn\\
43 & 314.01680 25.684672 & pow & 8.10$_{-2.51}^{+3.33}$ & $>$-1.35(V*) & agn\\
44 & 314.03226 25.693653 & apec & 1.42$_{-0.53}^{+0.72}$ & -2.50(V*) & star\\
45 & 314.04296 25.708472 & pow/apec & 7.75$_{-2.02}^{+2.61}$/5.46$_{-1.24}^{+1.88}$ & $>$-0.938/$>$-0.465(V*) & ?\\
46 & 314.06314 25.700192 & pow/apec & 1.32$_{-0.12}^{+0.13}$/0.977$_{-0.084}^{+0.079}$ & -0.288/0.0897(V*) & star\\
47 & 314.06771 25.673589 & apec & 4.64$_{-1.48}^{+1.50}$ & -3.16(V) & star\\
48 & 314.09986 25.658803 & pow/apec & 1.74$_{-0.78}^{+1.03}$/1.12$_{-0.63}^{+0.40}$ & $>$-0.435/$>$-0.146(V*) & ?\\
49 & 314.10914 25.736881 & pow & 1.63$_{-0.38}^{+0.45}$ & $>$0.0333(V*) & agn\\
50 & 314.03943 25.747156 & pow & 1.87$_{-0.10}^{+0.11}$ & 0.423(B) & agn\\
51 & 314.01804 25.758858 & pow & 3.77$_{-0.79}^{+0.97}$ & -0.0679(V*) & agn\\
52 & 313.95291 25.776819 & pow & 2.35$_{-0.14}^{+0.15}$ & 0.403(B) & agn\\
53 & 313.90073 25.751294 & apec & 1.09$_{-0.36}^{+0.47}$ & -2.12(V) & star\\
54 & 313.93571 25.858336 & pow & 2.23$_{-0.41}^{+0.46}$ & 0.0676(V*) & agn\\
55 & 313.77492 25.716167 & apec & 1.49$_{-0.39}^{+0.49}$ & -3.74(V) & star\\
56 & 313.95314 25.445281 & pow/apec & $<$1.00/$<$0.420 & $>$0.118/$>$0.448(V*) & ?\\
57 & 313.96553 25.421706 & pow & 1.34$_{-0.94}^{+1.79}$ & $>$0.000970(V*) & agn\\
58 & 313.97756 25.717731 & pow & 0.249$_{-0.228}^{+0.596}$ & -1.12(V*) & ?\\
59 & 313.99449 25.719272 & pow/apec & 6.10$_{-1.24}^{+1.50}$/4.73$_{-1.12}^{+2.59}$ & $>$-0.722/$>$-0.345(V*) & agn$^b$\\
60 & 313.90637 25.638078 & pow/apec & 2.17$_{-0.40}^{+0.46}$/1.38$_{-0.29}^{+0.33}$ & -1.43/-1.93(V) & star\\
61 & 313.90309 25.630814 & pow & 1.93$_{-0.30}^{+0.38}$ & -0.733(B) & ?\\
\caption{Analysis of the brightest serendipitous sources in the {\it XMM-Newton} FOV. The table shows the best source position, the statistically acceptable spectral model fit(s),
the fitted column density(s), the X-ray-to-optical flux ratio and the resulting identification. Typical XMM-Newton positional errors are 3$''$.
We reported the 1$\sigma$ error on column density.\\
a : Optical-to-X-ray flux ratio for the different fitting models. Where present, we used the V-band optical flux.
If not detected, we used the B-band optical flux. In case of an R-band detection, we obtained the V-band flux (see Appendix \ref{appa})\\
b : Identification of these sources is based on optical and hardness ratio tests.
}
\end{longtable}
\end{center}
\end{landscape}
\end{footnotesize}

\section{The Pulsar} \label{pulsar}

\begin{figure}
\epsscale{1.0}
\plotone{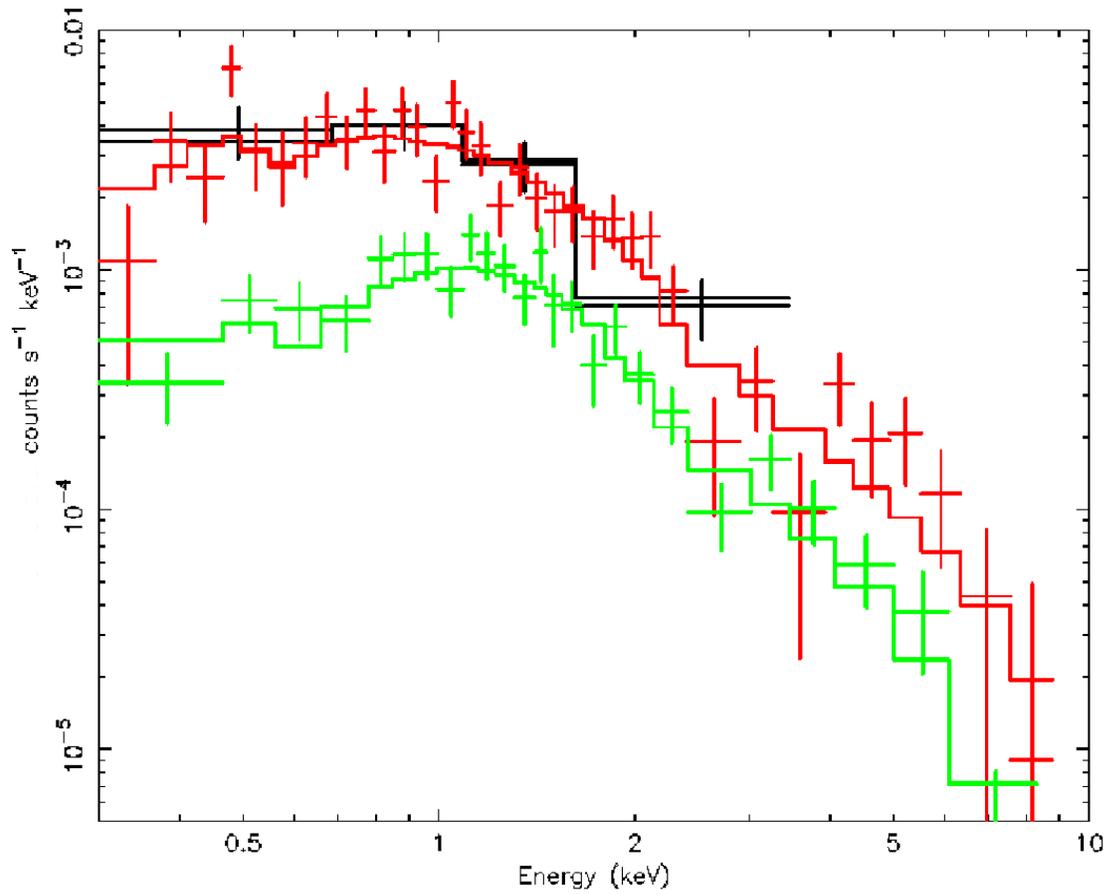}
\caption{{\it XMM-Newton} spectra of PSR J2055+2539. The pulsar PN spectrum from the first observation is in black,
the PN spectrum from the second observation is in red and the merged MOSs spectrum is in green.
The best-fit spectral model is shown.
\label{psr-spec}}
\end{figure}

In order to optimize the pulsar extraction radius, we
evaluated the brightness profile of the putative counterpart to maximize
both the signal-to-noise ratio (S/N) and the number of counts.
We found the best extraction radius to be 15$''$,
containing a total of 1020 and 636 counts in the 0.3-10 keV energy range
in the PN and the two MOSs detectors, respectively, with a background
contribution of $\sim$25\%. We generated ad-hoc response matrix
and effective area files using the SAS tools {\tt rmfgen} and {\tt arfgen}.
Due to the relatively low statistics and to optimize the fit, for the spectral analysis we added all the data from the MOS cameras using the HEAsoft tools
{\tt mathpha}, {\tt addarf} and {\tt addrmf} to sum spectra, effective area files and
response matrix, respectively. PN spectra were analyzed separately due to the different operation modes (Large Window and Full Frame)
requiring different response matrices.
The background region was extracted from a nearby source-free region common for every camera and on the same CCD of the source.
For the pulsar emission, we tried a simple power-law (pow), a simple blackbody (bb), a simple magnetized neutron star atmosphere (nsa),
the combination of a power-law and a blackbody (bb) as well as the combination of a power-law and a magnetized neutron star
atmosphere model ({\tt nsa} in XSPEC - assuming a neutron star with a radius of 13 km, mass of 1.4 M$_{\odot}$ and
a surface magnetic field of 10$^{13}$ G).
The simple power-law description gives an acceptable fit (reduced chi square $\chi^2_{red}$=1.05, 61 degrees of freedom, dof) with
a column density of N$_H$=2.18$\pm$0.26 $\times$ 10$^{21}$ cm$^{-2}$ and a photon index of $\Gamma$=2.36$\pm$0.14 (1$\sigma$ confidence errors).
The best-fit unabsorbed 0.3-10 keV flux is (3.43$\pm$0.27) $\times$ 10$^{-14}$ erg cm$^{-2}$ s$^{-1}$.
This flux is in agreement with the one reported in \citet{abd13} of 4.3$_{-2.8}^{+1.2}$ $\times$ 10$^{-14}$ erg cm$^{-2}$ s$^{-1}$.
Figure \ref{psr-spec} shows the spectra with the best-fit power-law model.
Neither the single component blackbody nor the nsa model give an acceptable fit ($\chi^2_{red}$=2.05 $\chi^2_{red}$=1.81 respectively, 61 dof)
and were therefore rejected.
Although the combined power-law plus thermal models give acceptable fits (both the bb and nsa give $\chi^2_{red}$=0.99, 59 dof), the thermal normalization
is compatible with zero at 2$\sigma$. Moreover, an F-test \citep{bev69} performed comparing the simple power-law with
the composite spectra does not point to a significant improvement by adding a thermal component.
We thus consider the single component non-thermal model to be the best for our source.

To search for X-ray pulsations, we used the most recent {\it Fermi} ephemeris~\footnote{\url{http://fermi.gsfc.nasa.gov/ssc/data/access/lat/ephems/}}.
As reported in Section \ref{obs}, we used only barycentered PN events from the 2013 {\it XMM-Newton} observation.
We decided to apply a photon weighting technique
similar to the one used for Fermi-LAT \citep[using {\it gtsrcprob}, ][]{ker11}, assigning to each photon a probability of coming from the
pulsar. This allows for a better background rejection and improves the sensitivity to pulsations, as discussed in \citet{mar14b}.
To this end, we further developed the Python tool reported in \citet{mar14b} (see Appendix \ref{appb}).
Considering only photons with a probability greater than 0.01 of coming from the pulsar, we used a weighted
Markov Chain Monte Carlo algorithm \citep[MCMC, see e.g.][]{wan13}
to test and refine the $\gamma$-ray ephemeris. Due to the shortness of the X-ray observation, we fixed the derivatives of frequency to ones of the $\gamma$-ray ephemeris.
We obtain a best-fitted frequency F = 3.129286$\pm$0.000003 s$^{-1}$, in agreement with the $\gamma$-ray one.
With six harmonics, the resulting H-value of 59.09 gives a tail probability of 1.25 $\times$ 10$^{-11}$. Even taking into account
the $\sim$ 50 trials, we found the X-ray pulsation with a confidence greater than 6$\sigma$, thus further confirming the identification of our
X-ray candidate with the $\gamma$-ray pulsar.
Unfortunately, the low statistics hamper a detailed energy-dependent timing analysi as was done in the case of the CTA-1 pulsar \citep{car10}.
This could have allowed us to unambiguously accept or exclude a possible low-energy (e.g. thermal) component of the spectral model.

\begin{figure}
\epsscale{0.5}
\plotone{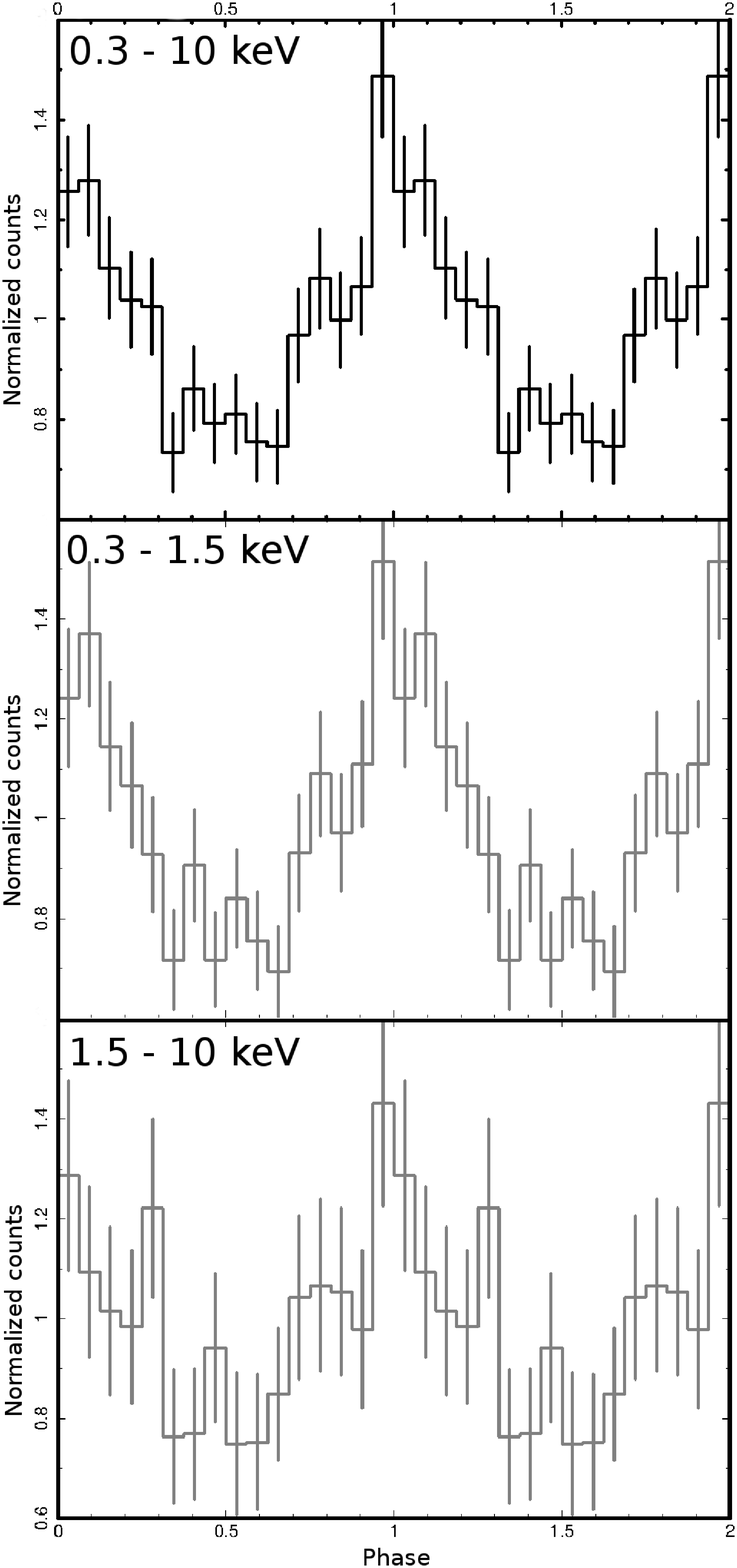}
\caption{{\it XMM-Newton} folded, weighted light curve of PSR J2055+2539. We normalized curves to have the mean of the distribution equal to 1.
The top curve is produced using the entire energy range,
the middle one using only photons with energies below 1.5 keV and the bottom one using photons with energies greater than 1.5 keV.
\label{psr-curve}}
\end{figure}

\section{Nebular emission} \label{nebula}

The most significant feature in the {\it XMM-Newton} FOV (Figure \ref{fig-field}) is a thin, straight, extended emission passing through half of the detectors
north to south and protruding from the position of J2055.
This feature is detected in each dataset, so that we exclude an instrumental origin.
Much less obvious is another elongated extended emission protruding from J2055 is visible south-east to north-west.
Unfortunately, this fainter secondary feature is heavily contaminated by bright point sources, hampering a classical analysis \citep[see e.g.][]{mar13}.

In order to better disentangle the nebular emission from the background and the numerous superimposed point-like sources,
we developed a new python-based tool to simulate and subtract their contribution from {\it XMM-Newton} data and images,
based on their positions and spectra (see Appendix \ref{appb}).
After running our tool, for each observation we created a new dataset that excludes photons based on the probability,  P,  that they come from
one of the considered sources (pulsar, serendipitous sources and background): using a Poisson distribution, each photon has a probability P of being excluded.
The remaining photons then come from either statistical fluctuations, or other sources not considered, such as our extended features.
We then based our analysis of extended sources on this reduced dataset, subsequently confirming our results on the original datasets.

\begin{figure}
\epsscale{1.0}
\plotone{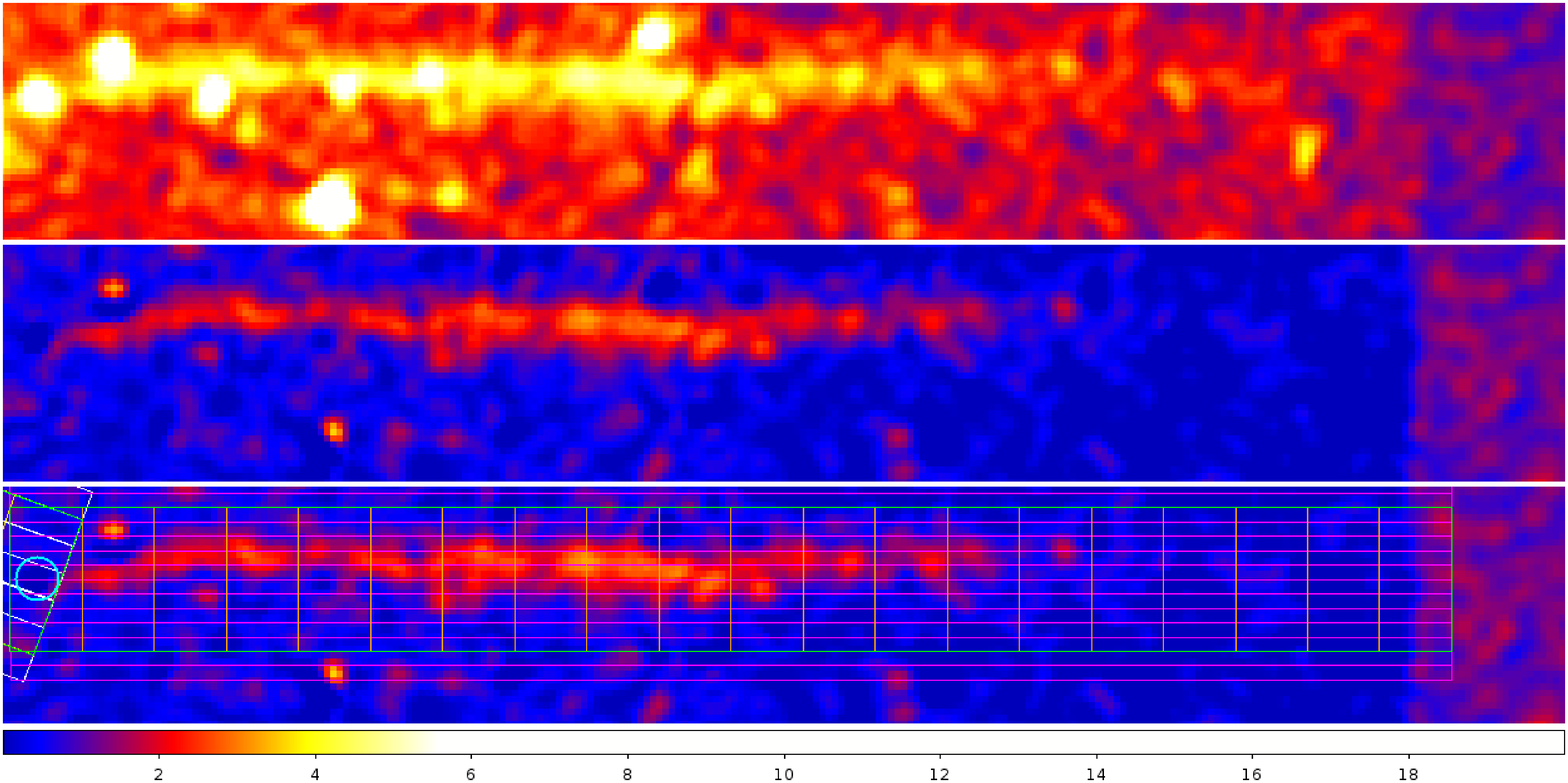}
\caption{Here, we show the sum of MOS2 {\it XMM-Newton} images. We applied a Gaussian smoothing with a kernel radius of 5".
Events are selected as described in Section \ref{obs}.
{\it Upper Panel:} This panel shows the main nebular feature as seen by {\it XMM-Newton}.
{\it Middle and Lower panels:} These panels show the main nebular feature with the point-like
sources and background removed (see Section \ref{nebula}). The lower panel also shows the regions used for the
spatial analysis in Subsection \ref{neb-ima}.
\label{ima-neb1}}
\end{figure}

\begin{figure}
\epsscale{1.0}
\plotone{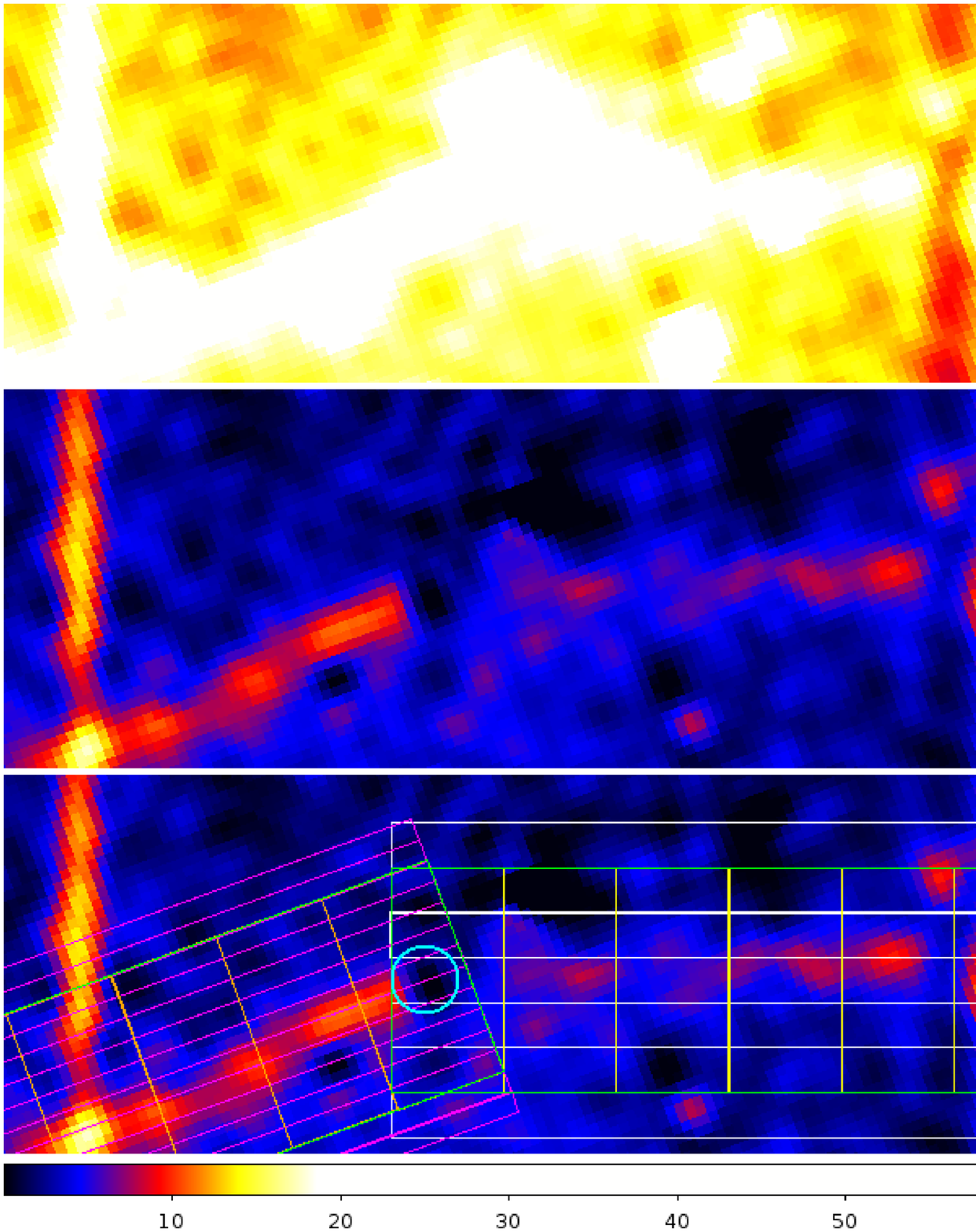}
\caption{Here, we show the sum of {\it XMM-Newton} images. We applied a Gaussian smoothing with a kernel radius of 5".
Events are selected as described in Section \ref{obs}.
{\it Upper Panel:} This panel shows the secondary nebular feature as seen by {\it XMM-Newton}.
{\it Middle and Lower panels:} These panels show the secondary nebular feature with the point-like
sources and background removed (see Section \ref{nebula}). The lower panel also shows the region used for the
spatial analysis in Subsection \ref{neb-ima}.
\label{ima-neb2}}
\end{figure}

\subsection{Spatial analysis} \label{neb-ima}

\begin{figure}
\epsscale{0.5}
\plotone{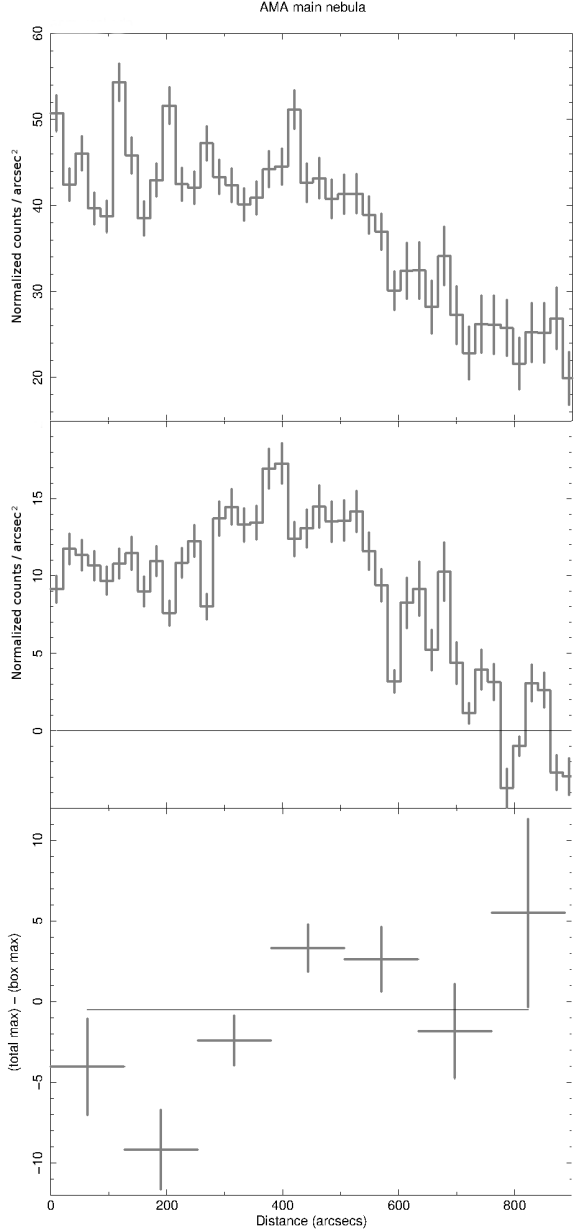}
\caption{Results from the spatial analysis along the main axis of the main nebula. Here, and in the following figures,
we plot the exposure-corrected counts versus the distance from the pulsar.
{\it Upper Panel:} Using the original {\it XMM-Newton} datasets.
{\it Middle Panel:} Subtracting the point-like sources and the background (see subsection \ref{neb-ima}).
{\it Lower Panel:} Difference between the best-fit maximum in each box and the total maximum, using a Lorentzian plus constant fit (see subsection \ref{neb-ima}).
\label{ima-1ama}}
\end{figure}

\begin{figure}
\epsscale{1.0}
\plotone{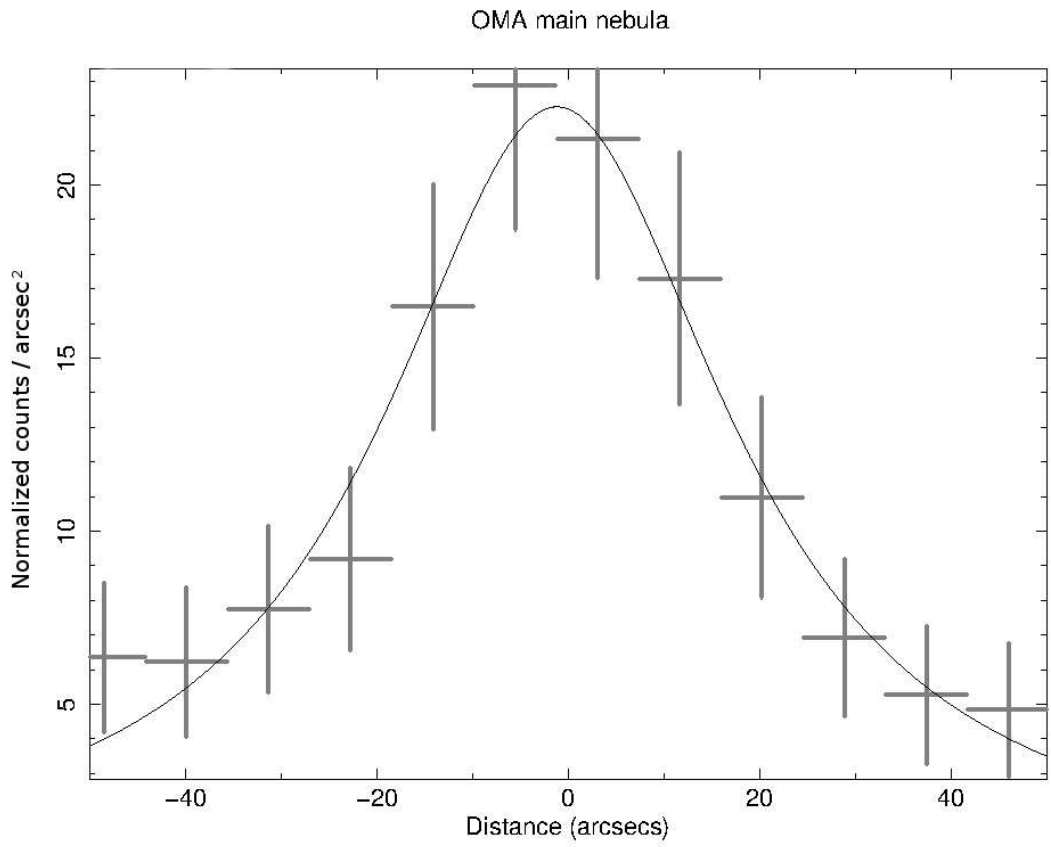}
\caption{Results from the spatial analysis orthogonal to the main axis of the main nebula; here, we subtracted the point-like sources and the background.
We reported the best-fit Lorentzian model.
\label{ima-1oma}}
\end{figure}

\begin{figure}
\epsscale{0.5}
\plotone{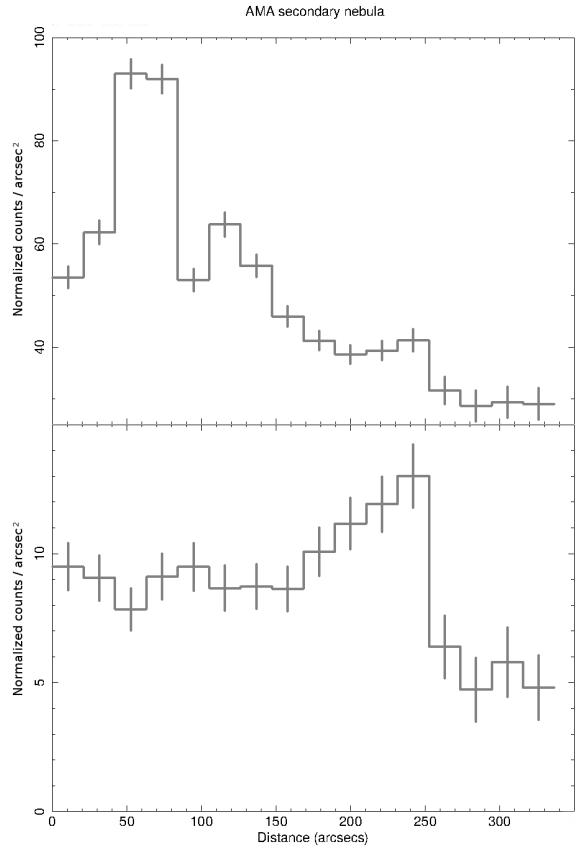}
\caption{Results from the spatial analysis along the main axis of the secondary nebula.
{\it Upper Panel:} Using the original {\it XMM-Newton} datasets.
{\it Lower Panel:} Subtracting the point-like sources and the background (see subsection \ref{neb-ima}).
\label{ima-2ama}}
\end{figure}

\begin{figure}
\epsscale{1.0}
\plotone{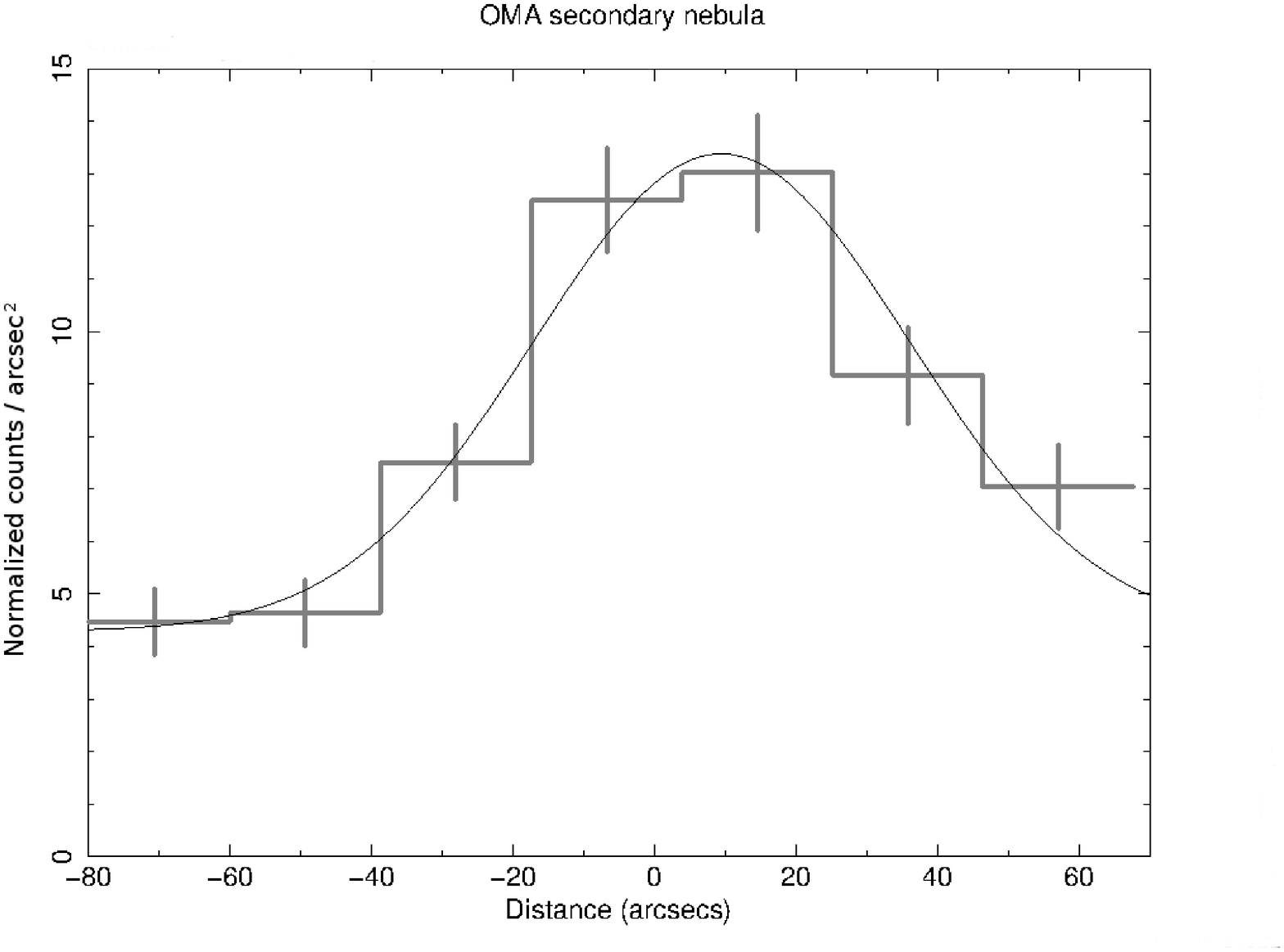}
\caption{Results from the spatial analysis orthogonal to the main axis of the secondary nebula; here, we subtracted the point-like sources and the background.
We reported the best-fit Gaussian model.
\label{ima-2oma}}
\end{figure}

For each camera, we produced brightness profiles of each nebula along the main axis (AMA) of the feature and orthogonal to the main axis (OMA).
At this point, we chose the main axis only through a visual evaluation, ending at the pulsar position.
In order to maximize the statistical significance of each bin, we carefully evaluated the bin size and the width of extraction
boxes, choosing a scale of 100"$\times$50" AMA and 1000"$\times$10" OMA for the main feature (showin in orange and magenta, respectively,
in Figure \ref{ima-neb1}) and 100"$\times$50" AMA and 400"$\times$20" OMA for the secondary feature (shown in yellow and white, respectively, in Figure \ref{ima-neb2}).
We extracted the number of counts in each box, summing the results of each dataset.
In order to take into account the spatial evolution of the exposures, as well as the lack of data from an entire camera at some points,
we also summed the exposure maps, produced through the SAS tool {\tt eexpmap}, and we built a graph of the exposure evolution AMA and OMA,
using the same boxes as before. This graph was then normalized at the mean value of the exposure. Finally, we divided the counts profile by the exposure profile
to obtain a normalized brightness profile, in normalized counts arcsec$^{-2}$.
Then, for each nebula we proceded iteratively by slightly changing the main axis in order to find the maximum of the OMA profile compatible with zero. Therefore, we will consider
this as the main axis of each nebula and its error as the propagated error from the maximum of the OMA profile.
The brightness profile for both the main feature are shown in Figure \ref{ima-1ama} and \ref{ima-1oma}
and for secondary feature are shown in Figure \ref{ima-2ama} and \ref{ima-2oma}.

First, we tested the presence of the fainter nebula.
To do this, we chose visually an ad-hoc area containing most of the nebular photons.
we considered different nearby regions with the same shape and on the same CCDs as the secondary nebula extraction area.
After considering the different exposure as well as instrumental effects, we obtain an expected number of counts from background
in the secondary nebula extraction region of $\sim$ 12600.
We estimated, conservatively, that $\sim$10\% of the total counts are due to residual photons from point-like sources superimposed on the nebula,
thus obtaining a total contribution of $\sim$ 800 counts. We detected a total of 15130 counts from the secondary nebular region
(after taking into account all the instrumental effects). The tail probability of obtaining 15130 counts from a Poisson distribution peaked at 13400 confirms the presence of this feature
at more than 7$\sigma$.
A similar calculation performed on the raw data, with 95\% of the counts from point sources removed using circular regions, gives
as well a nebular detection at more than 5$\sigma$.
We considered the possibility that such an excess comes from point-like sources not detected by the source detection algorithms described in Section \ref{field}.
These cannot realistically account for such a high number of additional counts:
by taking into account the upper limit on the detectable flux, we expect no more than $\sim$100 total counts from each of them, thus requiring a large number of unresolved sources.

We obtain the main axis of the main nebula to be at (-0.1$\pm$0.2)$^{\circ}$ with respect to the north-south axis, counterclockwise.
In the same reference, the secondary nebula axis is at (162.7$\pm$0.7)$^{\circ}$ (1$\sigma$ errors).

As shown in Figure \ref{ima-1ama}, along the main axis the main nebula is characterized by a slow increase in flux until $\sim$9', followed by a rapid drop until it
becomes undetectable at $\sim$12'. Orthogonal to the main axis (Figure \ref{ima-1oma}), the nebular profile is clearly peaked, following a Lorentzian-like 
distribution that fades symmetrically to the axis, $\sim$20" from it. We note that the OMA distribution is, at the limit, compatible with
the theoretical {\it XMM-Newton} PSF distribution ($\chi^2_{red}$=2.8, 11dof): this points to a thin (no more than $\sim$20"-thick), straight nebula
or a thinner, curved nebula.
To detect a possible main feature curvature  with respect to the main axis, we extracted counts from boxes of 170"$\times$125" AMA.
Each box was than divided into 15"$\times$125" sub-boxes, following the same procedure as before for the exposure, then
we fitted each box with a Gaussian. All the fits are statistically acceptable and the values (and errors)
of the maximum position are reported in Figure \ref{ima-1ama}.
The resulting curve is not compatible with a constant model ($\chi^2_{red}$=4.2, 6dof): the main nebula shows a curvature
in the clockwise direction at $\sim$3', then another in the counter-clockwise direction at $\sim$9'.

Along the main axis, the secondary nebula is characterized by an almost flat flux until $\sim$150", then slowly increasing until $\sim$250",
where it suddenly fades. Orthogonal to the main axis, it is peaked following a Gaussian-like distribution with a
sigma of 27"$\pm$5" (1$\sigma$ error). It does not follow the theoretical {\it XMM-Newton} PSF distribution ($\chi^2_{red}$=17.3, 6dof):
it is thicker (at least 50"-thick) than the main nebula. The low statistics prevent us from carrying out a more detailed spatial analysis.

Brightness profiles of the tails in the soft (0.3-1.5 keV) and hard (1.5-6 keV) energy bands show no statistically significant differences
from the total profiles.

\subsection{Spectral analysis} \label{neb-spec}

For the spectral analysis, we evaluated the best nebular extraction regions from the results of the spatial analysis.
For each nebula, we chose extraction boxes that ensure that the nebula is seen with more than a 3$\sigma$ significance.
This results in a box of 700"$\times$50" for the main feature and a box of 250"$\times$80" for the
secondary feature, oriented as their main axes. These boxes were also divided in two equal parts along the main axis in order to study the nebular spectral evolution.
For spectral analysis, we only took into account cameras that cover more than 50\% of the  considered region.
The normalizations of nebular spectra were left free to vary for each camera, due to the different FOV and coverage of the instruments.
We extracted background counts from regions with the same shape and angle, passing through the same CCDs as the source regions.
We chose these ad-hoc regions to be as near as possible to the source regions, but excluding the nebular features
and the extended feature west of the main nebula. In order to be conservative, we removed from the region 5$"$-radius circles centered at the position of point-like
sources, thus excluding possible residuals.
We followed the same procedure as in Section \ref{pulsar} to produce spectra, but we chose not to add them due to the higher statistics. Spectra were grouped
in order to have at least 150 counts per bin: owing to the high background, this allows us to have at least 25 source counts per bin in each camera.

A power-law model fits well both nebulae. The best-fit absorption columns of the nebulae and pulsar are compatible.
Then, under the hypothesis of association between the nebulae and the pulsar (see Section \ref{disc} for more details), we fitted all the spectra together, linking the column density.
In this way, we find the column density of the system to be (2.22$\pm$0.17) $\times$ 10$^{21}$ cm$^{-2}$ (90\% confidence error).
We note that a thermal bremsstrahlung model fits well both the nebulae, with column densities again in agreement with the one of the pulsar.
We found no sign of spectral evolution with distance from the pulsar for both the nebulae down to a variation of 0.1 and 0.3 (90\% confidence) in the photon index of the main and secondary nebula, respectively. See Table 2 for details on the fits.

\begin{table*}
\begin{center}
\begin{tabular}{ccccc}
\tableline\tableline
src & N$_H$ & $\Gamma$/kT & flux$^a$ & $\chi^2_{red}$/dof\\
 & {\tiny 10$^{21}$ cm$^{-2}$} & -/keV & {\tiny 10$^{-14}$ erg cm$^{-2}$ s$^{-1}$} & \\
\tableline
total & 2.22$\pm$0.17 & - & - & -\\ 
psr & 2.18$\pm$0.26 & 2.36$\pm$0.14 & 3.43$\pm$0.27 & 1.05/61\\
pwn1 & 2.25$\pm$0.23 & 1.82$\pm$0.08 & 28.7$\pm$1.9 & 1.19/249\\
pwn1-br & 1.68$\pm$0.16 & 5.41$_{-0.73}^{+0.92}$ & 23.5$\pm$1.8 & 1.15/249\\
pwn2 & 2.11$_{-0.59}^{+0.73}$ & 1.62$\pm$0.21 & 5.45$\pm$1.22 & 1.14/139\\
pwn2-br & 1.60$_{-0.40}^{+0.49}$ & 10.3$\pm$4.4 & 4.85$\pm$0.84 & 1.15/139\\
pwn1a & - & 1.75$\pm$0.12 & -$^{b}$ & 1.11/786\\
pwn1b & - & 1.82$\pm$0.11 & - & -\\
pwn2a & - & 1.50$\pm$0.33 & - & -\\
pwn2b & - & 1.77$\pm$0.20 & - & -\\
\tableline\tableline
\end{tabular}
\tablenotetext{a}{Unabsorbed 0.3-10 keV flux. Here, we report the flux of the instrument with the more complete view of the nebula
(respectively MOS 2 of the second observation and PN of the second observation for the main and secondary nebula).}
\tablenotetext{b}{The flux of the two parts of the nebulae is useless in order to describe them. For a complete spatial analysis see subsection \ref{neb-ima}.}
\caption{Results from spectral analysis of pulsar and nebulae. All the errors are at 1$\sigma$ confidence level.
Here, we report the column density obtained by fitting the pulsar and the two nebular spectra simultaneously (total).
We report the pulsar spectrum. We report the spectra of each of the nebulae (the main, 1, and the secondary, 2). Lastly, we report the contemporaneous
spectra of the two parts (nearer to pulsar, a, and furthest from to pulsar, b) of the two nebulae, with the column density linked.
}
\end{center}
\end{table*}

\section{Discussion} \label{disc}

\subsection{The pulsar}

In this paper we studied the X-ray counterpart of PSR J2055+2539.
While the majority of X-ray emitting isolated neutron stars \citep[NS, see e.g.][]{kas06} exhibit thermal emission coming from the whole NS surface,
no such component is required to fit our {\it XMM-Newton} spectra.
Using the blackbody model, for a NS radius of 10 km and a distance of 600 pc, the 3$\sigma$ upper limit on the surface temperature is 7.5 $\times$ 10$^5$ K.
This low temperature fits well the behavior of old pulsars (e.g. the J0357+3205 pulsar - Morla hereafter - upper limit temperature is half this value, \citet{mar13}): we note that the cooling
mechanism is highly dependent on the magnetic field of the pulsar, so that different magneto-thermal evolution paths are expected \citep[see e.g.][]{pon09}.
Taking into account the thermal radiation from hot spot(s), straight estimates of neutron star polar cap size based on a
simple dipole magnetic field geometry, give a polar cap radius R$_{PC}$ = R(R$\Omega$/c)$^{1/2}$, where R is the neutron star radius, $\Omega$
is the angular frequency, and c is the speed of light \citep{del05}. In the case of J2055 we obtain a value of 400 m.
In that case, the 3$\sigma$ upper limit on the surface temperature is 3.0 $\times$ 10$^6$ K, too high to rule out the existence of such a type of emission
for J2055. We note that the low-energetic J2055 is close to the death line for the production of e$^{+/-}$ (responsible for polar heating)
by curvature radiation photons, thus making this type of emission unlikely. A pulsar with comparable energetics, J0357+3205, shows reduced polar cap heating \citep{mar13}.
The non-thermal luminosity of J2055 is in agreement with the empirical relation between the X-ray luminosity of rotation-powered pulsars
and their spin-down luminosity \citep{pos02,kar08b}. J2055 also fits well the empirical relation between the X-ray luminosity (L$_x$) and $\gamma$-ray luminosity (L$_{\gamma}$) of radio-quiet pulsars, with a L$_x$/L$_{\gamma} = $1570$\pm$110 (1$\sigma$ confidence): the mean value for radio-quiet pulsars is $\sim$2500 \citep{mar15}.

We detected X-ray pulsations, thus further confirming the association of the X-ray source, with more than 6$\sigma$ significance. The shape of the pulse is consistent
with a sinusoid ($\chi^2_{red}$=1.17, 14 dof), although the statistics are too low for a complete characterization. The pulsed fraction is (25$\pm$3)\% (1$\sigma$ significance).
The pulsed fractions in 0.3-1.5 keV and  1.5-10 keV energy bands are consistent, (28$\pm$4)\% and (21$\pm$5)\% respectively.
A double-component pulsed emission usually show different shapes and peak position in the thermal and non-thermal components \citep[see e.g.][]{kas06},
thus the consistency among pulsed fractions calls for a single-component pulsed emission, making the pulsation consistent with a non-thermal origin.
We note that only $\sim$25 pulsars in the literature show non-thermal X-ray pulsation \citep[see e.g.][]{kui15}, fewer than half of these have a $\gamma$-ray counterpart and only three of them
are radio-quiet (Geminga, Morla and J1813-1246). This makes J2055 an interesting object to test models for high-energy emissions of pulsars.

\subsection{The nebulae}

Most of the analysis we carried out focused on the two elongated ragions of extended emission we detected inside the {\it XMM-Newton} FOV.
Both the sources have morphologies typical of PWNe or jets from pulsars, with an elongated shape, brighter at one edge and almost symmetrical
with respect to the main axis. Their spectra are non-thermal, best fitted by a power-law, thus confirming their association either with nebulae from pulsars or
a jet from an AGN.
An interpretation of the feature as an AGN jet can be safely discarded, owing to the angular extent of the feature: it
would imply an unrealistic physical size, unless the source is quite local (a huge 200 kpc long jet would imply an angular
scale distance of order 40 Mpc for the brighter feature and 150 for the fainter, assuming standard cosmological parameters), which would call for a rich multiwavelength
phenomenology (the host galaxy itself - with an angular scale well in excess of 1 arcmin - should be clearly resolved by
ground-based optical images we used). We thus refer to these features as pulsar nebulae.

The morphology of the nebulae, apparently protruding from J2055 and connected to the pulsar counterpart,
strongly argues for a physical association of the systems. Three other sources, 1a, 1b and 2 in Table 1,
can call for such a morphological association for the fainter nebula. Thanks to accurate spectral and multiwavelength analysis, we concluded that
source 1b and 2 are stars so that they cannot be associated to the features. The spectrum of source 1a, on the other hand, is non-thermal
and we found no optical counterpart that can be associated, up to a magnitude of 21. These characteristics point either to a pulsar or an AGN.
Taking into account the low number of known X-ray pulsars (to date, fewer than 200) in the entire {\it XMM-Newton} sky, the probability of finding a serendipitous pulsar within 1' of J2055 is negligible, thus confirming the association of both nebulae with J2055. Moreover, the best fitted
column density of source 1a is higher than the Galactic column density we found ((2.07$\pm$0.13) $\times10^{21}$ cm$^{-2}$),
thus pointing to an extragalatic object.
As further confirmation of the nebulae-pulsar association, their column density are in agreement, 2.18$\pm$0.26 $\times$ 10$^{21}$ cm$^{-2}$,
2.25$\pm$0.23 $\times$ 10$^{21}$ cm$^{-2}$ and 2.11$_{-0.59}^{+0.73}$ $\times$ 10$^{21}$ cm$^{-2}$ (90\% confidence) respectively for the pulsar, main nebula and secondary nebula.

The Galactic absorption column in the direction of the pulsar, predicted by \citet{dic90} at $\sim$450 pc and based on the H$_{\hbox{\scriptsize I}}$ distribution (1.2 $\times$ 10$^{21}$ cm$^{-2}$),
is lower than the best-fit value for the pulsar-nebulae system ((2.22$\pm$0.17) $\times10^{21}$ cm$^{-2}$), pointing to a lower limit for the pulsar distance
of $\sim$450 pc. This result is in broad agreement with the value estimated by scaling the $\gamma$-ray flux of the pulsar \citep{saz10}:
the method hinges upon the observed correlation between the intrinsic $\gamma$-ray luminosity and spin-down power of
the pulsar \citep{saz10,abd13}, assuming a beam correction factor of one for the $\gamma$-ray emission cone of all the pulsars \citep{wat09}.
By applying this relation, we have a $\gamma$-ray efficiency of 0.04, an X-ray efficiency of 3 $\times$ 10$^{-4}$ and a distance of 600 pc.
We obtain an upper limit of $\sim$750 pc by requiring the $\gamma$-ray efficiency to be less than 1.
Another important piece of information could come from the serendipitous stars we found in the field: their distance estimates
from optical analyses could provide limits on the distance of the pulsar. Future multi-filter optical observations will provide this limit.

We found two nebulae associated with J2055, with angular dimensions of 12' $\times$ 20" and 250" $\times$ 30", respectively, for the brightest and faintest.
The observed extensions of the nebulae, at a distance of 600 pc,
would correspond to physical dimensions of $\sim$ (2.1 $\times$ 0.05) pc and $\sim$ (0.7 $\times$ 0.09) pc respectively (assuming no inclination with respect to the plane of the sky).
The luminosities of the nebulae in the 0.3-10 keV energy range
(assuming d = 600 pc) are 1.2 $\times$ 10$^{31}$ erg s$^{-1}$ and 2.4 $\times$ 10$^{30}$ erg s$^{-1}$, corresponding
to fractions of 2.4 $\times$ 10$^{-3}$ and 3.0 $\times$ 10$^{-4}$ of the pulsar spin-down luminosity.
A few elongated tails of X-ray emission associated to rotation-powered pulsars have been discovered in the past
\citep{gae04,mcg06,bec06,kar08b}. Although for our pulsar we have no information
about the proper motion, the bow-shock PWN scenario would seem the most natural explanation for one of the nebulae.
Luminosity values are fully compatible with that measured for other synchrotron nebulae, for which the pulsar channel into their tails 10$^{-2}$ to 10$^{-4}$ of
their rotational energy loss.
In the case of synchrotron emission tails, if we assume the optimistic maximum Lorentz factor of $\sim$ 10$^8$ for electron acceleration in such a low-energetic pulsar magnetosphere,
and the high value of the ambient magnetic field of $\sim$ 50 $\mu$G \citep[following considerations in][]{del11},
it is possible to estimate the synchrotron cooling time of the
emitting electrons as $\tau_{sync}$ $\sim$ 100 (B/50$\mu$G)$^{-3/2}$ (E/1keV)$^{-1/2}$ yr. Coupling this value with the estimated physical length of
the feature yields an estimate of the bulk flow speed of the emitting particles of $\sim$ 20,000 and 3,000 km s$^{-1}$, assuming no inclination
with respect to the plane of the sky. The first value is only marginally consistent with results in literature, while the second one is fully consistent \citep{kar08b}.
Taking into account the energetics, both the nebulae are at least marginally consistent with a classical synchrotron nebula explanation.
For classical synchrotron nebulae, we expect a relatively bright diffuse emission surrounding the pulsar, where the emission from the
wind termination shock is brightest. The model of \citet{gae06} predicts a low-scale termination shock of $\sim$ 0.6$"$ fo J2055 (at 600 pc),
thus not resolved by {\it XMM-Newton}, even in the case of typical ambient
densities (0.01 atoms cm$^{-3}$) and pulsar velocities (some hundreds km s$^{-1}$). In that case, part of the flux
we assign to the pulsar comes instead from the termination shock \citep{kar08a}.
Along their main axis, the brightness profile of both the nebulae are consistent with an almost flat behavior, with a sudden decrease at the end.
This is acceptable for synchrotron emission nebulae: in fact, the
pulsar wind is more energetic in the surrounding of the pulsar, where the loss of energy through synchrotron emission is higher,
decreasing in flux and energy along the pulsar trail. Taking also into account an inhomogeneous medium and/or magnetic field,
this behavior is expected, and observed for many objects in the literature \citep[see e.g.][]{kar08b}.
Such a synchrotron cooling of the particles injected at the termination shock induces a significant softening
of the emission spectrum as a function of the distance from the pulsar in bow-shock PWNe. Taking into account the upper limit variation of the photon index we found,
for the brightest tail of J2055 we do not have the predicted spectral variation.
Classical theories of ram pressure particle confinement in pulsar tails are hardly put to the
test by parsec-long nebulae, that would require much higher efficiencies
than in all the other cases. The tightness of the main feature of J2055, if confirmed to be a synchrotron
nebula from its shape and the pulsar motion, cannot be accounted for by any model in the literature.
We conclude that the low energetics of the pulsar, the lack of any spatial-spectral variability
and the tightness of the brighter nebula disfavor the classical synchrotron emission nebula.
This model can explain the characteristics of the secondary feature.

If we consider that the main nebular emission comes from thermal bremsstrahlung, we can assume an
optimistic cone angle of the tail of 1$^{\circ}$. Taking into account the spectral best fit values and following \citet{mar13},
this implies an interstellar medium temperature of $\precsim$ 50,000 K, depending on the inclination of the nebula.
The pulsar velocity along the plane of the sky would be $\sim$ 2300 km s$^{-1}$, one of the highest values for a pulsar.
The high temperature of the pre-shock is consistent with that of the hot phase of the ISM, which fills a large fraction
of the Galaxy \citep{bla00}. The variation of the symmetry with respect to the main axis of the nebula
with the distance from the pulsar can be explained in terms of a clumpy distribution of the interstellar medium.
The long expected cooling time ($\sim$ 10$^7$ yr) also explains the lack of spatial-spectral variation of the nebula.
In any case, we need the lack of emission surrounding the pulsar for the electrons must be heated by the ions, heated by the
pre-shock flow. We have no indication about the interstellar medium density, that would depend on the inclination of the
system, but taking into account values similar to the ones of PSR J0357+32, we expect that within $\succsim$ 50" of the pulsar
there should be no nebular emission. The brightness profile along the main axis clearly rules out this model both for the 
main and the secondary nebulae.

Both scenarios we considered need
the proper motion of the pulsar to be aligned with the main axis of the nebula. The presence of two
different tail-like features implies that at least one of them cannot be explained in terms of classical
nebular models.
A magnificent example of a long, collimated X-ray nebular feature, misaligned with the pulsar
proper motion is the Guitar Nebula (the radio-loud PSR B2224+65, located at 1 kpc, \citet{hui07}).
The Guitar feature is much shorter ($\sim$2' long) and larger than the J2055 one, with the similar nebula-
pulsar flux ratio of $\sim$6. The main feature of J2055 is thus compatible with a Guitar-like feature from a
nearer object ($\sim$150 pc rescaling Guitar feature). This feature is usually explained in terms of a
collimated ballistic jet, even if more complex explanations have been developed \citep{ban08}.
These models predict a spectral change across the feature, excluded from {\it XMM-Newton}
observations in the case of J2055.
Recently, \citet{pav14a,pav14b} found a long helical jet-like structure protruding from the energetic pulsar IGR J11014-6103.
Two different features protrude from the point-like object, one shorter (1.2')
and wider probably aligned with the proper motion, also radio-emitting, and one 5.5'-long forming
an angle of $\sim$ 104$^{\circ}$ with the shorter feature axis. The jet scenario is the most probable, mainly due
to the helicoidal shape of the nebula, revealed by the Chandra observation. While similar to the J2055
feature in length, the fluxes of the IGR object and its two main features are comparable. With
J2055 we are facing a much more collimated nebula protruding from a much less energetic pulsar, theoretically difficult to account.
In order to confirm the helical pattern of the main nebula and proceed with theoretical modeling, as well as to confirm the misalignment of the proper motion, high-resolution
images of the nebula should be taken at different epochs.

We recently obtained two new X-ray observations of the J2055 system, taken by {\it Chandra}. The system will be observed during a long,
uninterrupted observation, allowing for a full characterization of the tails with a much better spatial resolution than {\it XMM-Newton}.
With such an observation, we will also improve the results of our script on {\it XMM-Newton} data.
After two years, {\it Chandra} will re-observe the J2055 system in a short observation. This will allow us to detect the proper motion
down to $\sim$ 350 km s$^{-1}$, for a distance of 600 pc and a null inclination with respect to the plane of the sky.
Such information will be crucial to understand the origin of the two nebular features of PSR J2055+2539.

\section{Conclusions} \label{conc}

We analyzed the {\it XMM-Newton} observations of the radio-quiet $\gamma$-ray PSR J2055+2539.
We confirmed the X-ray counterpart association of the pulsar. The X-ray emission is non-thermal, with
an X-ray efficiency of $\nu_x$ = L$_x$/$\dot{E}$ $\sim$ 3 $\times$ 10$^{-4}$ (for a distance of 600 pc), typical of X-ray pulsars.
The $\gamma$-to-X-ray luminosity ratio is L$_x$/L$_{\gamma}$ $\sim$ 1600, typical of radio-quiet pulsars.
The lack of cooling thermal emission is due to the old age of the pulsar, while the lack of hot spot thermal emission
can be due to its low energetics.
We detected non-thermal X-ray pulsations, following a sinusoidal profile with a pulsed fraction of $\sim$ 25\%.
Taking into account considerations on the $\gamma$-ray efficiency of the pulsar and on its X-ray spectrum,
we can infer a pulsar distance ranging from 450 pc to 750 pc.

We developed a new script to simulate the spatial distribution of {\it XMM-Newton} counts from input point-like and extended sources
Using this, we found and analyzed two different nebular features associated to J2055. Both nebulae have an elongated profile, protruding from the pulsar and almost symmetrical to
a main axis. The main axis of the two nebulae are separated by an angle of $\sim$ 163$^{\circ}$.
The main, brighter feature is 12'-long, $<$20"-thick (corresponding to 2.1 $\times$ 0.05 pc at a 600 pc distance), characterized by
an asymmetry with respect to the main axis evolving with the distance from the pulsar, possibly forming helical pattern.
The secondary feature is 250" $\times$ 30" (that corresponds to 0.7 $\times$ 0.09 pc at a 600 pc distance).
Both nebulae present an almost flat brightness profile, with a sudden decrease at the end.
The nebulae can be fitted either by a power-law model or a thermal bremsstrahlung model.
Their efficiencies (L$_x$/$\dot{E}$) are respectively 2.4 $\times$ 10$^{-3}$ and 3.0 $\times$ 10$^{-4}$ for the main and the secondary nebula.
A plausible interpretation of the brighter nebula is in terms of a collimated ballistic jet, mainly due to its thickness and shape.
The secondary nebula is most probably a classical synchrotron-emitting tail.
Future {\it Chandra} observations will allow us to better study the shape of the nebulae and to find the proper motion of the pulsar,
thus allowing us to test our interpretations.
Finally, we found and studied a bright, serendipitous galaxy cluster with a central AGN.

\appendix

\section{Appendix A: field analysis} \label{appa}

Based on the study of spectra and possible optical counterparts, we can classify serendipitous
sources as AGN or candidate stars, allowing us to constrain the pulsar distance. Indeed, after selecting
candidate AGN in the FOV, it is possible to measure from their spectra the total Galactic
column density in the direction of J2055.

For this analysis, we decided to use only brighter sources, those from which it is possible to extract and fit a spectrum:
we therefore require that they have an absorbed flux greater than 5 $\times$ $10^{-15}$ erg cm$^{-2}$ s$^{-1}$,
with a 5$\sigma$ detection and with at least 150 {\it XMM-Newton} net, total counts.
We also considered all the sources detected in the nebular regions.
As described in Section \ref{field}, we found 57 serendipitous bright X-ray sources in the field of view.
In order to classify them, we performed a spectral analysis, as done in \citet{mar14b}.
All the sources, with the exception of source 1, are compatible with a point-like distribution.
As described in Appendix \ref{appa}, a deeper analysis of source 1 revealed it as the superimposition of two or three different sources.
Extended sources are discussed elsewhere.
As a first method of classification, we used their spectra. Based on \citet{nov09}, at mid-Galactic latitude we expect
that $\sim$75\% of serendipitous sources with the considered flux are AGN and the remaining are stars.
X-ray emission from AGN is best described by a power-law, while X-ray emission from stellar coronae by single or double {\tt apec} (emission spectrum from collisionally-ionized diffuse
gas). For sources located inside extended emissions, we extrapolated the spectrum and flux of the extended source and froze it into the spectral
fit.
Out of these 58 sources, 30 can be realistically fitted only be a power-law and 5 only by apec(s); for the remaining 23 objects, this type of analysis is
not conclusive (see Table 1).

Since the count statistics of some of the selected X-ray sources are too low to discriminate the spectral
model, we performed a qualitative spectral analysis using the count rate (CR). 
The CR is measured in three energy ranges: soft ($0.3-1$ keV), medium ($1-2$ keV) and hard ($2-10$ keV).
We then computed the two different hardness ratios:

\begin{eqnarray*}
 HR12=[CR(1-2)-CR(0.3-1)]/[CR(1-2)+CR(0.3-1)]\\
 HR23=[CR(2-10)-CR(1-2)]/[CR(2-10)+CR(1-2)] 
\end{eqnarray*}

\begin{figure}
\epsscale{1.0}
\plotone{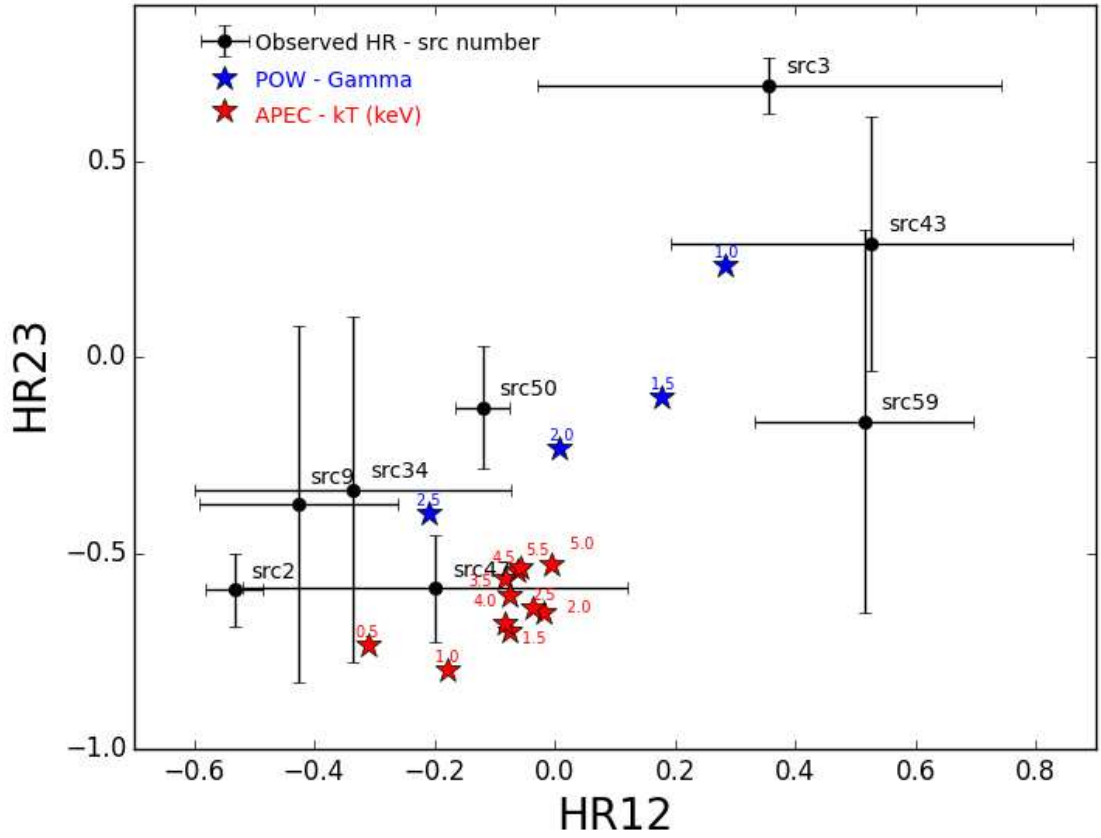}
\caption{Distribution of HR12 vs. HR23 of a sample of selected X-ray sources. Error bars are reported at 1$\sigma$. Blue stars indicate
the expected HR12 vs. HR23 computed for power-law spectra with $\Gamma$ from 1.5 and 2.5. Red stars indicate the expected HR12 vs.
HR23 computed for apec spectra with kT from 0.5 to 5.5 keV. Each spectral model is computed using the average interstellar
medium absorption given by \citet{dic90}.
\label{fig-hr}}
\end{figure}

Large/small HR12 values points to a large/small absorption, a large/small HR23 indicates a hard/soft spectrum.
The values of the hardness ratios are compared to a matrix of simulated values for apec
(kT ranging from $0.5$ to $5.5\,keV$) and power-law ($\Gamma$ ranging from $1.5$ to $2.5$) models. 
Each spectral model is computed using the average interstellar medium absorption given by \citet{dic90}. 
We extracted and simulated count rates from the PN camera
of the second observation and the sum of MOSs cameras, then we combined them.
These values are plotted in Figure \ref{fig-hr} together with a sample of the 
measured HR ratios. From the position of the latter with respect to the former,  an interpolation is possible over the two models and the values of the model parameters.
Although this method proved best for the case of
the highly-absorbed serendipitous sources in the field of PSR J1813-1246 \citep{mar14b}, for J2055 field this method did not improve the spectral results already
obtained for most of the sources. Only in the case of source 2 do we have an indication of a star-like spectrum and for sources 3, 43 and 59 of AGN-like.

Another common way to confirm X-ray classification of sources is based on multiwavelength analysis: the X-to-optical flux ratio is a good indicator
of the nature of X-ray emitters. According to \citet{lap06}, AGN have typical logarithms of X-to-optical flux ratios higher than
-0.15(+0.2), while stars lower than +1.0(+0.65) in the V(B) band. Inside each X-ray source error box, we looked for association with optical sources from the NOMAD catalogue
\citep{zac05}, considering the V-band magnitude as reference or B-band where necessary. Alternatively, we used the R-band magnitude:
V magnitude was then evaluated from R using the following method.
We considered the best-fitting X-ray models to estimate the column densities for each source. In case of multiple fitting models, we repeated the following for each model.
We de-absorbed R magnitudes by using coefficients from \citet{pre95,car89},
obtaining the R-band absolute magnitude. Then, assuming a thermal spectrum and/or a power-law spectrum,
based on the X-ray model, we used the on-line NICMOS tool \footnote{\url{http://www.stsci.edu/hst/nicmos/tools/conversion\_form.html}} to evaluate the V-band absolute magnitude.
Few of our AGN-like objects have optical counterparts inside their error
box that could be due to spurious matches. In order to estimate the number of spurious matches,
we used the relation from \citet{sev05,nov09}. This yielded a probability
of chance coincidence of 21\%, wich means that, at our limiting magnitudes, contamination effects
cannot be ignored. Each of the objects revealed as star-like from spectral or HR analysis has an optical counterpart that agrees with the expected
X-ray-to-optical flux ratio. Table 1 reports the associated optical counterparts and expected
upper limits for AGN.

Finally, we combined results from spectral and multiwavelength analysis to classify our serendipitous sources, with the restrictions reported above.
All the sources with spectral model or HR compatible with a star coronal emission and with a star-like X-to-optical flux ratio are considered as stars: these are 13 objects.
All the sources with spectral model or HR compatible with an AGN emission and with a AGN-like X-to-optical flux ratio are considered as stars: these are 24 objects. 21 sources cannot be uniquely classified.

\section{Appendix B: a script for X-ray photons weighting} \label{appb}

\begin{figure}
\epsscale{1.0}
\plotone{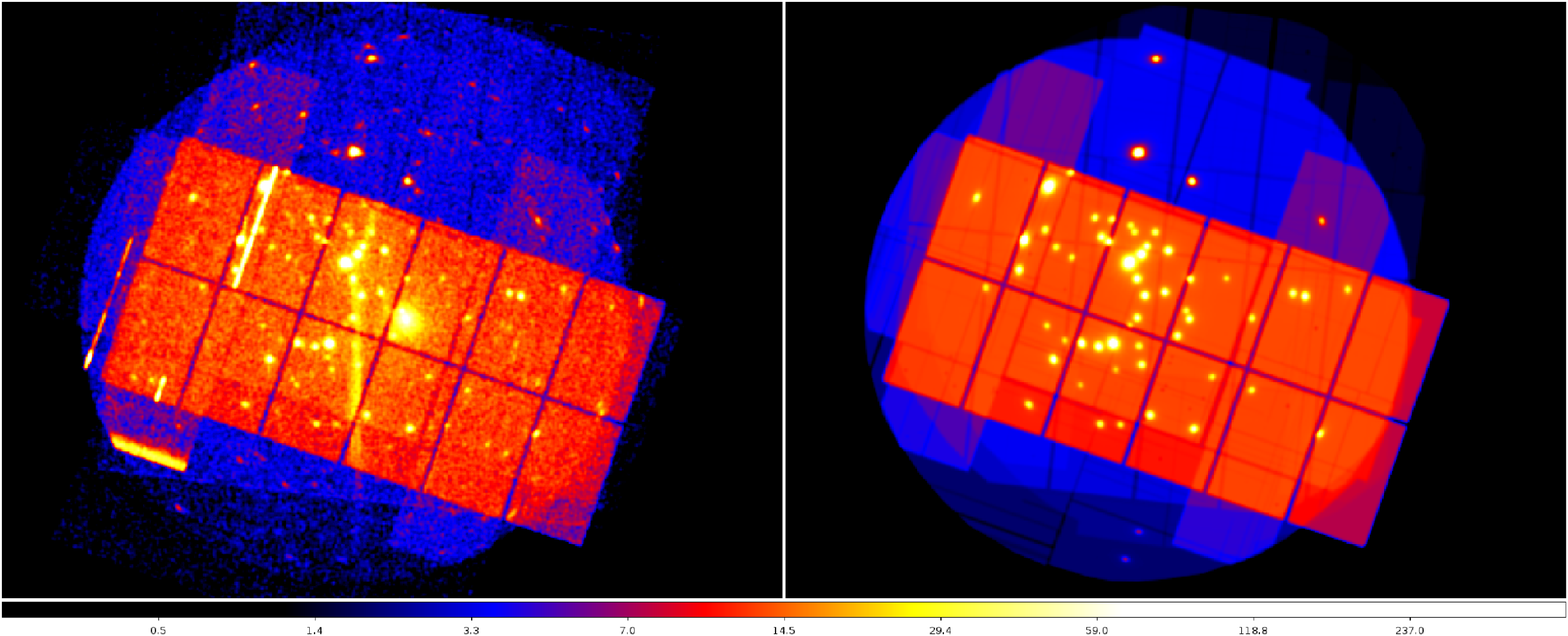}
\caption{{\it Left Panel:} sum of all the 0.3-10 keV {\it XMM-Newton} counts images, represented in logarithmic scale.
Here, we did not apply all the filters described in Section \ref{obs} to remove bright columns;
 {\it Right Panel:} sum of all the 0.3-10 keV {\it XMM-Newton} images simulated by our script, represented in logarithmic scale.
The asymmetry of the PSF is apparent for sources with high off-axis angles
and hints of spikes are visible for the brightest sources; examples of the differences between the two figures are showed in Figures \ref{ima-neb1}, \ref{ima-neb2}}.
\label{sim1}
\end{figure}

We developed a script to simulate the spatial distribution of photons from input point-like and extended sources.
After this simulation, by setting the value of the variable `other', it is possible either i) to search for new sources using the reduced dataset without photons from considered sources
(other=yes) or ii) to extract photons from a single source to perform high-level spectral and timing analysis (other=no).
We note that this type of approach is particularly well suited to combine the high spatial resolution of {\it Chandra-ACIS} observatory with the high spectral and timing
resolutions of {\it XMM-Newton}. The first satellite can provide the number and exact position of point-like sources as well as the flux and shape of nebular
features: these can be used as inputs of the script to perform high-level {\it XMM-Newton} analysis.
In this paper, first we used the script to clean out the nebular features from point-like sources reported in Table 1, see Figure \ref{sim1}.
Then, we used again the script to extract photons from the pulsar to perform the timing analysis.

We used python v2.7.5 language to build the script. This python script recalls standard python modules - {\tt os, sys, math, shutil, multiprocessing, subprocess} -
as well as modules dedicated to scientific analysis - {\tt xspec, numpy, pyfits}.
This script has a number of variable input parameters:\\
- the type of analysis: to search for new sources or to extract photons from a source (`other' variable);\\ 
- the energy range and energy resolution to be used, in eV;\\
- the spatial accuracy to be used for point-like PSF reconstruction; if one of the sources is extended, it should be set also the extended PSF accuracy;\\
- the names and positions of input and output files.\\
The script takes a list of inputs from an ASCII file:\\
- the position of each source, both in ecliptic and physical coordinates;\\
- the spectral model and parameters of each source, readable by xspec module;\\
- the source extraction circle radius from which each spectrum of point-like source is extracted;\\
- the spectrum of background, readable by the xspec module;\\
- the extraction circle radius of background spectrum.\\
Finally, as input it needs some FITS files:\\
- event file, filtered by the user for instrumental effects and energy;\\
- image file constructed from the event file;\\
- exposure file constructed from the event and image files;\\
- calibration file (for {\it XMM-Newton}, the CCF file);\\
- the spectrum, background spectrum, response matrix (rmf) and effective area (arf) file from each source and from the background;\\
- if extended sources are loaded, for each of them an image of the theoretical spatial shape of the nebula (before the PSF application), possibly with different normalizations of each pixel.\\
As output, it produces:\\
- for each input source, an ASCII file containing all the commands ran by the script;\\
- for each input source and the background, a column is added to the fits file: this show the probability (0$<$P$<$1) of the photon to come from that source;\\
- for each input source, the image of the simulation of that source, see Figure \ref{sim1};\\
- an image containing all the simulated sources, an image with residuals and an image containing the deviation of simulated map from counts map, in sigmas.
Due to the high computing time needed, each source is treated separately on a different CPU and the results are merged only at the end, to compute
probabilities and create images.

The first part of the script produces the mean count rate over the entire observation of each source, in each energy band.
Here, the spectrum, background spectrum, rmf and arf are loaded using the {\tt pyxspec} library and the input spectral
model is assigned (but not fitted). Based on the spectral model interpolated with rmf and arf, the script computes the count rate.
We note that these values are corrected for PSF tails and exposure map.

The second part of the script constructs the PSF of each source, in each energy band.
We used the last theoretical model for {\it XMM-Newton} PSF, as reported in \citet{rea11} and following calibration documents.
The normalization of the PSF function is set in order to obtain the correct number of expected count rates, as calculated in the first part of the script.
In case of an extended source, this is treated as a number of point-like sources with positions and normalizations following the input model image.
We note that, in order to reduce the computing time, a different, lower-resolution, spatial grid can be set for extended sources.
The use of this complicated, spatial and energy-dependent, PSF function rather than the simple King function previously adopted resulted in a significant
improvement in the positional accuracy, as well as spurious sources identification for automated {\it XMM-Newton} searches of serendipitous sources.
This change in the PSF becomes more important for bright sources in the field, allowing for
a better disentangling photons from each source, as well as the detection of nearby faint sources.
In the case of the J2055 field, the adoption of this newer PSF resulted in the clear identification of the
two, or possibly three, different sources responsible for the emission from source 1 in Table 1.

The third part of the script computes the number of observed counts, from each source in each energy band, in each pixel of a spatial grid, with a step
set by the user. It is based on the theoretical PSF function and values obtained in the second step. Finally, the obtained matrix is multiplied for the
input exposure map to obtain the number of expected counts.

The fourth part of the script creates output images. The selected spatial grid is collapsed to agree with the input counts map resolution and results from all the energy bands
are summed. For each source, a simulated image is written and they are all summed to write the total simulated sky.
The script also creates a residual image, subtracting the simulated map (S) from the counts map (C), and a significance map defined for each i bin as SI$_i$=$\|$O$_i$-S$_i\|$/$\sqrt{O_i}$
where O$_i$!=0 and SI$_i$=0 otherwise.

The fifth part of the script computes the probabilities. The script assigns to each photon in the FITS file a list of predicted counts values for each source,
based on the pixel of the spatial grid where the photon falls. If `other'=yes, it also takes into account the observed number of counts in that pixel.
If higher than predicted, it adds the value (number of observed counts)-(number of predicted counts) to the list, otherwise it adds a zero.
Finally, the sum of the values for each photon is normalized to 1 and the obtained list of probabilities is added to the input FITS file.
We note that the `other=yes' mode can be used only to search for new sources but not for the analysis of an already-detected source:
in fact, this would induce a systematic error due to the wrong characterization of the statistics. On the other hand, accurate analysis
of the residuals would allow one to find low-statistic point-like sources or extended sources that can be added to the input list
to iteratively improve the results of the script.
In order to extract photons from the source of interest, only sources $\sim$ 7.5' away from it should be considered. {\it XMM-Newton} theoretical
PSF is null further than 5' from the source, so that only sources that affect this circle are necessary.

In order to validate this method, we applied it to several different pulsar fields: we used the weighted MCMC analysis presented in Section \ref{pulsar}
to compare the H-value resulting from weighted and unweighted tests. We obtained improvements of H-value ranging from 10 to 50\%, with better results for
pulsars with multi-component spectra, superimposed on nebular and/or point-like sources.
In the case of the already-published J1813-1246 \citep{mar14b}, a simpler version of this script increased the H-value from 11123 to 12092;
the use of this new method further increased it to 12432. In the case of J2055, the H-value increased from 52 to 66 (with the same number of trials).
In the case of Geminga, we used some of the public {\it XMM-Newton} observations to detect the non-thermal pulsation, in the 2-10 keV band.
Using our new script, we found an increase in the H-value from 120 to 220, almost doubling it. Tests on other pulsars are in progress.
We also considered the distribution of residuals of the considered point-like sources. For each of the $\sim$ 50 object considered, we found residuals of
no more than a few percent of the original fluxes, thus confirming the validity of the method.

The case of source 1 (see Table 1) is representative of the capacity of this script. That object was revealed both by the SAS tool {\tt edetect\_chain}
and the CIAO tool {\tt wavdetect} as a single point-like emission. A brightness profile analysis does not reveal significant deviations from the theoretical
{\it XMM-Newton} PSF. A spectral analysis gives instead a peculiar result: this source cannot be fitted by a single-model spectrum but requires a double-component
spectrum ({\tt apec+apec} or {\tt apec+power-law}). Our script allowed for a more accurate analysis of the source.
Taking into account a single point-like object as source 1, we obtain residuals of more than
40\% of source counts (while the other $\sim$50 sources give only a few percentage points),
distributed in pixels south-east to north-west from the best source 1 position; we also obtain an overestimation of counts in pixel
south-west to nort-east, so that the total number of simulated counts is in agreement with the observed ones. We note that the nearby nebula should not
significantly contribute to the source counts, due to its faintness. Then, we tried a model with two different sources: the positions were iteratively
refined in order to minimize the residuals. We found that two sources separated by $\sim$8$''$ describe adequately the `source 1 system', with lower residuals of $\sim$15\%. The brightest one has a flux
$\sim$10 times larger than the weakest and its spectrum is best described by a power-law while the other one spectrum is best described by an {\tt apec};
spectral parameters and optical counterparts allow us to associate them with an AGN and a star, respectively. We note that a three-sources system,
forming an equilateral triangle with a side of $\sim$ 10$''$ would lower the residuals to $\sim$3\%.

Although this program is much more evolved than the script we used for the analysis of PSR J1813-1246 \citep{mar14b}, it still needs further technical
improvement to reduce the high computing time. A future version of this script could also be more user-friendly: a simplified version of the analysis
could be run using the energy bands and associated count rates for each of the objects reported in the 3XMM catalog of serendipitous {\it XMM-Newton}
sources \footnote{\url{http://xmmssc.irap.omp.eu/Catalogue/3XMM-DR5/3XMM\_DR5.html}}. Future implementations could also allow the user to analyze data from
X-ray observatories different than {\it XMM-Newton}, e.g. {\it NuSTAR} \citep{har13}, by simply changing the PSF model and some variables.

\acknowledgments

This work was supported by the ASI-INAF contract I/037/12/0,
art.22 L.240/2010 for the project $''$Calibrazione ed Analisi del satellite NuSTAR$"$.

{\it Facilities:} \facility{XMM}.

\clearpage

\end{document}